\documentclass{raa_onecolumn}
\usepackage{graphicx,times}
\usepackage{natbib}
\usepackage{amssymb,amsmath}
\usepackage{booktabs}
\usepackage{hyperref}
\usepackage{colortbl}
\bibpunct{(}{)}{;}{a}{}{,}

\newcommand{\sect}[1]{Section~\,\ref{#1}}

\newcommand{\fig}[1]{Figure~\,\ref{#1}}
\newcommand{\figs}[1]{Figures~\,\ref{#1}}
\newcommand{\eqn}[1]{Equation~\,(\ref{#1})}
\newcommand{\eqns}[1]{Equations~\,(\ref{#1})}

\begin{document}

 \title{Application of a magnetic-field-induced transition in Fe~{\sc{x}} to solar and stellar coronal magnetic field measurements}

 \volnopage{ {\bf 2022} Vol.\ {\bf XX} No. {\bf XX}, 000--000}
   \setcounter{page}{1}

  \author{Yajie Chen$^{1,2}$, Wenxian Li$^3$, Hui Tian$^1$, Xianyong Bai$^3$, Roger Hutton$^4$, Tomas Brage$^5$ } 

   \institute{
$^1$School of Earth and Space Sciences, Peking University, Beijing 100871, China; {huitian@pku.edu.cn}\\
$^2$Max Planck Institute for Solar System Research, G\"{o}ttingen 37077, Germany\\
$^3$National Astronomical Observatories, Chinese Academy of Sciences, Beijing 100101, China; {wxli@nao.cas.cn} \\
$^4$Department of Astronomy, Beijing Normal University, Beijing 100875, China \\
$^5$Division of Mathematical Physics, Lund University, Post Office Box 118, SE-221 00 Lund, Sweden \\
\vs \no
   {\small Received 2022 XXX; accepted 2022 XXX}
}

\abstract{Magnetic fields play a key role in driving a broad range of dynamic phenomena in the atmospheres of the Sun and other stars. Routine and accurate measurements of the magnetic fields at all the atmospheric layers are of critical importance to understand these magnetic activities, but in the solar and stellar coronae {such a measurement is still a challenge due to the weak field strength and the high temperature}.
{Recently, a magnetic-field-induced transition (MIT) of Fe~{\sc{x}} at 257.26 {\AA} has been proposed for the magnetic field measurements in the solar and stellar coronae.}
In this review, we present an overview of recent progresses in the application of this method in astrophysics.
We start by introducing the theory underlying the MIT method and reviewing the existing atomic data critical for the spectral modeling of Fe~{\sc{x}} lines. We also discuss the laboratory measurements that verify the potential capability of the MIT technique as a probe for diagnosing the plasma magnetic fields.
We then continue by investigating the suitability and accuracy of solar and stellar coronal magnetic field measurements based on the MIT method through forward modeling. Furthermore, we discuss the application of the MIT method to the existing spectroscopic observations obtained by the Extreme-ultraviolet Imaging Spectrometer onboard \textit{Hinode}. This novel technique provides a possible {way for routine measurements of the magnetic fields in}
the solar and stellar coronae, but still requires further efforts to {improve its accuracy}.
Finally, the challenges and prospects for future research on this topic are discussed.
}

   \authorrunning{{\it Chen et al.}: Research in Astronomy and Astrophysics}            
  \titlerunning{{\it MIT in Fe~{\sc{x}}}: Research in Astronomy and Astrophysics}  
   \maketitle

%

\section{Introduction}\label{sect:intro}

Magnetic fields are essential in the understanding {of} many phenomena and activities in the atmospheres of the Sun and late-type stars.
{As an example}, the evolution of magnetic field structures drives solar and stellar flares and coronal mass ejections (CMEs) \citep[e.g.,][]{2020RAA....20..165L,Tan2020,2021NatAs...5..697V,2019NatAs...3..742A,LuHP2022,ChenHechao2022}, which may result in significant interplanetary disturbances \citep[e.g.,][]{2022SciA....8I9743H} and severe space weather \citep[e.g.,][]{2020RAA....20...36C,2020RAA....20...23S}.
The magnetic fields are also the energy source of many types of small-scale heating events, such as Ellerman bombs \citep[e.g.,][]{2021RAA....21..229C}, ultraviolet bursts \citep[e.g.,][]{2014Sci...346C.315P}, jets \citep[e.g.,][]{Tian2018} and coronal bright points \citep[e.g.,][]{2019LRSP...16....2M,2020RAA....20..138N}, which may contribute to the heating of the solar and stellar coronae \citep[e.g.,][]{Parker1983,Parker1988,2021A&A...656L...7C}.
Despite the importance of magnetic fields in the solar atmosphere, accurate and routine field measurements are only available only for the photosphere {but not for} the upper atmosphere \citep[e.g.,][]{Wiegelmann2014}.
Measurements of the stellar coronal magnetic fields are even more challenging and {only limited results are achieved}.

Methods based on the Zeeman effect have been used to obtain the solar photospheric magnetic field from spectropolarimetric observations of magnetic-sensitive photospheric lines \citep[e.g.,][]{Iniesta2016,2020RAA....20...45H}, and to measure the average magnetic flux density on the stellar surfaces \citep[e.g.,][]{2012LRSP....9....1R}.
The Zeeman-Doppler imaging technique has also been developed to derive large-scale magnetic field distribution on the surfaces of some fast-rotating stars \citep[e.g.,][]{1989A&A...225..456S,2015ApJ...805..169R}.
However, coronal magnetic field measurements based on Zeeman effect are extremely difficult since field strength is several orders of magnitudes smaller, while the temperature is several orders of magnitudes higher, than in the photosphere. This results in the Zeeman splitting being negligible compared to the line broadening.
\citet{Lin2000,Lin2004} have attempted to measure the coronal magnetic field in off-limb active regions from spectropolarimetric observations of the near-infrared Fe~{\sc{xiii}} 10747 {\AA} line. 
However, their observations were integrated over roughly one hour to obtain sufficient signal-to-noise ratios, which makes it difficult to investigate the temporal evolution of the coronal magnetic field.
In addition, the spectropolarimetric diagnostics of the coronal magnetic field using infrared lines are only limited to active regions.
The spectropolarimetric technique has in some cases been applied to measurements of the magnetic field of full coronal-loop systems using chromospheric lines emitted from cool plasma in the corona loops \citep[e.g.,][]{Schmieder2014,Kuridze2019}.

Since oscillations and waves are prevalent in the solar atmosphere \citep[e.g.,][]{2021RAA....21..179J,2021RAA....21..126W,2020RAA....20..117F,2020RAA....20....6W},
some physical parameters, such as the magnetic field, can be inferred from their dispersive properties from coronal observations \citep[e.g.,][]{2020ARA&A..58..441N,2020SSRv..216..136L,2021SSRv..217...76B}.
This method, termed coronal seismology or magnetoseismology, was proposed by \citet{Uchida1970} and \citet{Roberts1984}.
\citet{Aschwanden1999} and \citet{Nakariakov1999} found decaying transverse oscillations of coronal loops triggered by a flare. Subsequently, \citet{Nakariakov2001} calculated the magnetic field strength based on the dispersive properties of the oscillations and obtained a value of $\sim$10 G.
A similar method has also been applied to magnetic field measurements in other coronal structures such as coronal steamers \citep{Chen2011}.
However, these decaying transverse oscillations are often triggered by solar flares or CMEs, so they can only be observed occasionally.
Another limitation is that these early studies can only provide an average value or 1D distribution of the magnetic field strength \citep[e.g.,][]{2015A&A...581A.137C,Li2018}.
Similarly, quasi-periodic pulsations (QPPs) on other stars are often observed during stellar flares \citep[e.g.,][]{1974A&A....32..337R,1986ApJ...305..363L,2006A&A...458..921W}, and some studies have attempted to derive the stellar coronal magnetic field strength in the flare regions from these observations \citep[e.g.,][]{2005A&A...436.1041M,2009ApJ...697L.153P}.
Nevertheless, the physical mechanisms of stellar QPPs are still under debate \citep[e.g.,][]{2018SSRv..214...45M}, which impedes its application to the measurement of the coronal magnetic field of stars.
There are also ubiquitous waves and decayless oscillations in the solar corona \citep[e.g.,][]{Tomczyk2007,Tian2012,Wang2012,Morton2015}. By applying the magnetoseismology technique to these waves and oscillations, we may obtain 2D distributions or temporal evolution of the coronal magnetic field \citep[e.g.,][]{Long2017,Magyar2018}.
Recently, \citet{Yang2020a,Yang2020b} applied magnetoseismology to the pervasive propagating transverse waves in the corona and obtained the first global coronal magnetic field map.
However, their method can only provide the plane-of-sky component of the magnetic field above the solar limb.

Radio imaging observations from radioheliographs such as the Nobeyama Radioheliograph \citep[NORH, ][]{NORH1,NORH2}, the Expanded Owens Valley Solar Array \citep[EOVSA, ][]{EOVSA}, and the  Mingantu Spectral Radioheliograph \citep[MUSER, ][]{MUSER,2021RAA....21..284Z} can also be used to infer information about the coronal magnetic field.
Through spectral fitting, radio spectral observations often give magnetic field strengths from tens to hundreds of Gauss in solar active regions  \citep[e.g.,][]{Akhmedov1982,Akhmedov1986,Wang2015,Miyawaki2016,Iwai2013}.
Recently, \citet{Anfinogentov2019} reported that the magnetic field strength at the base of the corona in an active region could reach 4000 Gauss.
Radio observations have also been applied to magnetic field measurements in flaring structures \citep[e.g.,][]{Gary2018,Chen2020,Zhu2022,2016RAA....16...82T}.
For example, \citet{Fleishman2020} observed obvious decay of the magnetic field strength during a flare, indicating that the eruption is triggered by magnetic reconnection.
In addition, radio observations have been used to infer the coronal magnetic field strength for active regions or flare regions on some other stars \citep[e.g.,][]{1981ApJ...250..284G,2002ARA&A..40..217G}. 
But it is a challenge to determine the radio emission mechanisms crucial for magnetic field measurements from radio observations \citep{2022RAA....22g2001T}.

Another approach to determine the coronal magnetic field structures is by using magnetic field extrapolation from photospheric magnetograms taken from observations in an active region \citep[e.g.,][]{Sun2012,Aschwanden2013,Wang2015nc,Chifu2017,Wiegelmann2021,Zhu2018,Zhu2022} or the whole corona \citep[e.g.,][]{Schatten1969,Aly1984,Tadesse2014}.
In addition, the combination of extreme ultraviolet (EUV) or infrared observations and magnetic-field models
\citep[e.g.,][]{2008ApJ...680.1496L,2009AnGeo..27.2771L,2017ApJ...838...69L,2018ApJ...856...21C,2022RAA....22g5007Z} or magnetohydrodynamic (MHD) models
\citep[e.g.,][]{2011ApJ...731L...1D,2013SoPh..288..617R,2016FrASS...3....8G,2021ApJ...912..141Z,2019ApJ...883...55Z, Jiang2022} can also aid in the understanding of the magnetic field structures in the corona.
Some previous studies determined the global stellar magnetic field structures above the surfaces through force-free extrapolation from magnetograms obtained from the Zeeman-Doppler imaging technique \citep[e.g.,][]{2002MNRAS.333..339J,2006Sci...311..633D,2014MNRAS.437.3202J}.
However, these models are often based on many assumptions, such as the force-free field or magnetohydrostatic equilibrium, which are not always valid in the solar and stellar atmosphere \citep[e.g.,][]{Peter2015_extraplation}.

Recently, three of the present authors (WL,RH and TB) recognized that magnetic fields could also be measured by using unexpected transitions induced by external magnetic fields, so called magnetic-field induced transitions (MIT). In certain circumstances, described below, these transitions are enhanced due to accidental close degeneracy between levels of short and long lifetimes. This led to a systematic search for candidates, by investigating occurrences of close degeneracy between short- and long-lived levels in different atomic spectra. The most promising candidate existed in Fe~{\sc{x}}, which is discussed in this review. This discovery led to the first theoretical investigations of these transitions in the  Modern Physics institute of Fudan University in Shanghai~\citet{Li2015,Li2016}. Here it was also proposed that this exotic transition in Fe~{\sc{x}} is a candidate for a new coronal magnetic field diagnostic technique since its intensity is significantly affected by the strength of the local magnetic field, external to the observed ions.
This MIT method has since then received strong attention from the solar physics community. In an early paper spectra from the SO82-B spectrograph on board SKYLAB was used in a first attempt to determine the splitting between the two close-to-degenerate levels~\citep{Judge2016}. It has since then been applied to the spectral observations taken by the EUV Imaging Spectrometer \citep[EIS,][]{Culhane2007} onboard \textit{Hinode} \citep{Hinode} for the measurements of coronal magnetic fields in active regions at the limb and on the disk \citep{Si2020,Landi2020a,Brooks2021}.
Based on forward modeling with 3D MHD models of solar and stellar coronae, \citet{Chen2021a,Chen2021b} and \citet{Liu2022} have demonstrated that the MIT method could provide coronal magnetic field measurements in solar active regions and Sun-like stars with a strong surface magnetic flux.
%

In this review, we first briefly introduce the theoretical background of the MIT method, then summarize recent progresses in the application of the MIT method to solar and stellar coronal magnetic field measurements, with a focus on the validation of the technique through the approach of forward modeling and attempts of coronal magnetic field measurements from actual solar coronal observations.
We also discuss various sources of uncertainty that should be reduced before implementation of the technique in routine measurements of the solar and stellar coronal magnetic field. 

\section{History of magnetic-field-induced transitions} \label{sect:mit}

The influence of magnetic field on atomic energy levels, known as the Zeeman effect, has been widely used for the measurement of magnetic field in various astronomical objects since its first attempt for the sunspot by \citet{1908ApJ....28..315H}.
The magnetic interaction also breaks the symmetry of an atomic system and allows the mixing between atomic states with the same magnetic quantum number and parity.
This will in turn introduce a new decay channel, namely MIT, and thereby “unexpected” lines to appear in spectra and lifetimes of the long-lived state to be shortened.
The impact of the external magnetic field on the ion is usually very small due to the relative weakness of this field in comparison to the strong internal field of the ions (hundreds or even thousands of tesla). 
Therefore, the effect usually only contributes to very weak lines or long-lived metastable states.
For example, the impact of the magnetic field on the lifetime of a long-lived level has been investigated by \cite{1967Phy....33..278F}. It was labeled Zeeman quenching, in order to determine the intrinsic lifetime of the metastable state.
\cite{1992PhRvL..69.1042B} and \cite{1993PhRvA..47..890A} observed a decrease in the lifetime of the metastable Be$^-$(2s2p$^2$ $\mathrm{^4P_{3/2}}$) and He$^-$(1s2s2p $\mathrm{^4P_{5/2}}$) states, respectively, utilizing a Heavy-Ion Storage Ring.
Subsequently, \cite{1997HyInt.108..291M} and \cite{2005PhRvA..72b0501S} measured the decay rates from the Xe$^+$(5p$^4$5d $\mathrm{^4D_{7/2}}$) state at different magnetic field strengths using the Ion Storage Ring and obtained the lifetime of $\mathrm{^4D_{7/2}}$ by a nonlinear extrapolation to field-free conditions.

The first MIT line in spectra was observed by \cite{2003PhRvL..90w5003B} for the $\mathrm{2p^53s~^3P^o_0}$ –  $\mathrm{2p^6~^1S_0}$ transition in Ne-like Ar using the EBIT-II electron beam ion trap (EBIT) in the Lawrence Livermore National Laboratory. 
In the presence of an external magnetic field, the usually strictly forbidden one-photon transition channel $\mathrm{^3P^o_0}$ –  $\mathrm{^1S_0}$ is opened due to the mixing of the upper state $\mathrm{^3P^o_0}$ with $\mathrm{2p^53s~^{3,1}P^o_1, M=0}$ states, which can decay to the ground state with an intercombination transition and an allowed electric dipole (E1) transition, respectively.
\cite{2003PhRvL..90w5003B} illustrated that MIT could be used for magnetic field strength diagnostics in high-temperature plasmas.
\cite{2013PhRvA..88a3416L} performed a theoretical investigation for $\mathrm{2p^53s~^3P^o_{0,2}}$ –  $\mathrm{2p^6~^1S_0}$ MIT rates in Ne-like ions between Mg~{\sc{iii}} and Zn~{\sc{xxi}} and demonstrated that it is important to include both perturber states $\mathrm{^1P^o_1}$ and $\mathrm{^3P^o_1}$ in order to produce accurate transition rates.
\cite{2004RScI...75.3720B} and \cite{2016ApJ...817...67B} further measured the MIT lines in Ne-like Ar and Fe, respectively, using EBIT-II.
The measured $\mathrm{^3P^o_0}$ –  $\mathrm{^1S_0}$ MIT rate for Ne-like Fe is in agreement with the theoretical prediction by \cite{2013PhRvA..88a3416L}.
\cite{2013PhRvA..88b2513G} carried out theoretical calculations for the $\mathrm{2s2p~^3P^o_{0}}$ –  $\mathrm{2s2p~^1S_0}$ MIT transition in Be-like isoelectronic sequence between Z = 5 and 92 and suggested that the effect of the magnetic field needs to be evaluated in order to properly measure the E1M1 two-photon decay rate in a storage ring.
Unfortunately, the MIT in Ne-like and Be-like ions is for rather strong fields (several Tesla) and is therefore not feasible for use in solar and stellar atmospheres. However, if the quantum states end up very close to each other in energy, the perturbation by the external field will be enhanced.
   
\begin{figure}
\centering
\includegraphics[width=10.0cm, angle=0]{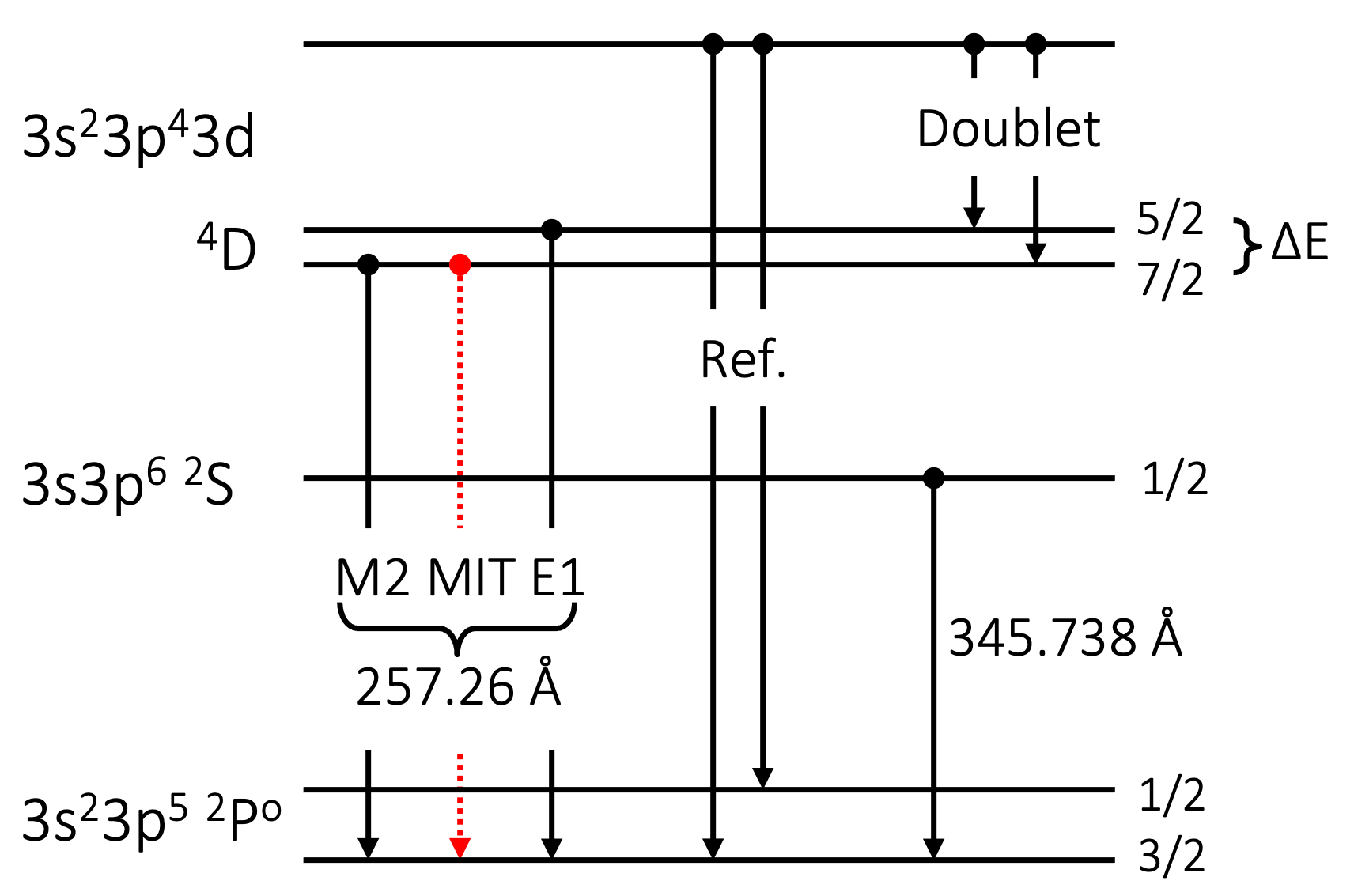}
\caption{Schematic energy diagram and decay channels for the levels of the Fe~{\sc{x}} ion that are relevant to the MIT method.
Reference lines (Ref.) and doublet are listed in Table \ref{tab:lines}. The fine-structure energy separation $\Delta E$ = $E(\mathrm{^4D_{5/2}})$ - $E(\mathrm{^4D_{7/2}})$.}
\label{fig:levels}
\end{figure}
    
A close, accidental degeneracy of two quantum states was observed in Fe~{\sc{x}} by \cite{Li2015} (a partial energy level diagram of Fe~{\sc{x}} is shown in \fig{fig:levels}).
The close degeneracy occurs between the $\mathrm{3p^43d~^4D_{7/2}}$ level, which in a field-free space only decays with a slow magnetic-quadrupole (M2) transition to the ground level $\mathrm{3p^5~^2P^o_{3/2}}$, and the $\mathrm{3p^43d~^4D_{5/2}}$ level, which decays with a faster E1 allowed transition to the ground state (see \fig{fig:levels}). 
The influence of the external magnetic field opens an allowed E1-MIT transition from $\mathrm{^4D_{7/2}}$ to the ground state through mixing with the $\mathrm{^4D_{5/2}}$ level.
It is worth noting that, due to the close degeneracy, it is far beyond the resolution power of any available EUV spectrometer and therefore the two lines will appear as a blend in the spectra, being at 257.26 \AA. 
It is fortunate that this close degeneracy occurs in Fe~{\sc{x}}, since this ion has a high abundance in astrophysical plasmas including the solar corona, and the line at 257 \AA~is one of the strong Fe~{\sc{x}} lines in solar coronal spectra observed by \textit{Hinode}/EIS.
\cite{Li2015} and \cite{Li_2021} presented systematic theoretical studies for the atomic properties needed to determine the dependence of the MIT rate on the strength of the magnetic field and proposed that the MIT in Fe~{\sc{x}} can be used to measure the magnetic field strengths in the solar corona.
More recently, \cite{Xu2022} measured the MIT of Fe~{\sc{x}}~at different magnetic fields using the Shanghai High-temperature Superconducting Electron Beam Ion Trap (Shanghai-Htsc EBIT).


\section{Theoretical method, Atomic data, and Laboratory Measurement} \label{sect:data}

\subsection{General theory}

The Hamiltonian of an atom with zero nuclear spin under the influence of an external homogeneous magnetic field {B} can be written as
\begin{equation}\label{fig:H}
H = H_{fs} +  H_{m} = H_{fs} + ( {\bf N}^{(1)}  +  \Delta {\bf N}^{(1)}) \cdot {\bf B}
\end{equation}
where $H_{fs}$ is the relativistic fine-structure Hamiltonian and $H_m$ is the interaction Hamiltonian with the external magnetic field.
The tensor operator ${N}^{(1)}$ represents the coupling of the electrons with the field, 
and $\Delta {N}^{(1)}$ is the Schwinger QED correction. Further details of the operators can be found in \cite{Cheng1985}.
In this case, $M$ remains the only good quantum number (apart from parity), and the M-dependent atomic state wave function $| M \rangle$ can be written as an expansion
\begin{eqnarray}
\label{eq:AW}
| M \rangle = \sum_{\varGamma J} d_{\varGamma J} | \varGamma J M \rangle
\end{eqnarray}
with mixing coefficients associated with the magnetic-field perturbation $d_{\varGamma J}$ obtained by solving the corresponding eigenvalue problem or through first-order perturbation theory,
\begin{eqnarray}
\label{eq:mixing}
d_{\varGamma J} = \frac{\langle\varGamma J M|H_m|\varGamma_0 J_0 M_0\rangle}{E(\varGamma_0 J_0)-E(\varGamma J)}
\end{eqnarray}
where $|\varGamma_0 J_0 M_0\rangle$ represents the reference atomic state.
The electric dipole transition probability for a MIT from an initial state $|M'\rangle$ to a final state $|M\rangle$ is given by
\begin{eqnarray}
\begin{aligned}
\label{eq:MIT-fs}
{ A(M' \rightarrow M) = \frac{2.02613 \times 10^{18}}{\lambda^3}\sum_q \bigg | \sum_{\Gamma J}\sum_{\Gamma' J'}d_{\Gamma J}d'_{\Gamma' J'}} (-1)^{J-M}  
                                                 \begin{pmatrix}
                                                  J  & 1 & J'    \\
                                                 -M  & q & M'
                                              \end{pmatrix}
\langle\Gamma J || {\bf P}^{(1)} ||\Gamma' J' \rangle \bigg |^2.
\end{aligned}
\end{eqnarray}
where $A(M' \rightarrow M)$ is in s$^{-1}$ and $\lambda$ is the wavelength of the transition in \AA. {The average transition rate $A_{\mathrm{MIT}}$ can be obtained by $A_{\mathrm{MIT}}=\frac{\sum_MA(M)}{2J+1}$.}

The general theory can be applied to the MIT rates in Fe~{\sc{x}}. 
The reference state $\mathrm{3p^43d~^4D_{7/2}}$ ($|7/2\rangle$ for simplicity), under the influence of an external magnetic field, can approximately be expressed as
\begin{eqnarray}
\label{eq:fex}
|``7/2"M\rangle \approx d_{7/2} |7/2M\rangle + d_{5/2} |5/2M\rangle
\end{eqnarray}
The field-independent atomic state functions $|7/2M\rangle$ and $|5/2M\rangle$ can be calculated using the multiconfiguration Dirac-Hartree-Fock (MCDHF, \cite{MCDHF}) approach and the field-dependent mixing coefficients $d_{7/2}$ and $d_{5/2}$, as well as the MIT rates as a function of magnetic field strengths can be obtained using the GRASP2018 module HFSZEEMAN95 \citep{GRASP2018,Li2020}.

\begin{figure}
\centering
\includegraphics[width=10.0cm, angle=0]{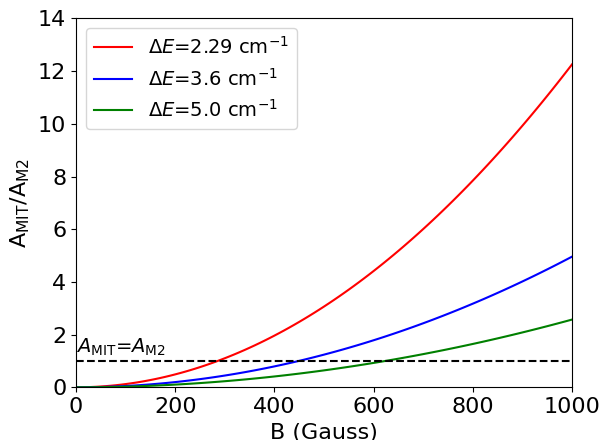}
\caption{$A_\mathrm{MIT}$/$A_\mathrm{M2}$ ratios as a function of magnetic field strengths for different energy separations $\Delta E$.}
\label{fig:mit-m2}
\end{figure}

\subsection{$\Delta E$} \label{sect:delta_E}
\begin{table*}
\label{tab:lines}
\caption{Fe X lines relevant to the MIT method. {The transition probabilities from \cite{Li_2021} are shown in the last column. Note that the rates of 257 \AA~ lines are given at B = 0 Gauss.}}
\centering
\begin{tabular}{ccccccccccccccccccccccccc}
\hline\midrule
&& i-j  &&  $\lambda$~({\AA})&&   Lower level                 && Upper level && {$A$ (s$^{-1}$)}                               \\  
\hline
MIT lines && 1-5  &&  257.259           &&   3s$^2$3p$^5$ $^2$P$^\mathrm{o}_{3/2}$ && 3s$^2$3p$^4$3d $^4$D$_{5/2}$ && {6.077e+06} \\  
 && 1-4  &&  257.261           &&   3s$^2$3p$^5$ $^2$P$^\mathrm{o}_{3/2}$ && 3s$^2$3p$^4$3d $^4$D$_{7/2}$  && {5.748e+01} \\  
\hline
&& 1-30 &&  174.531           &&   3s$^2$3p$^5$ $^2$P$^\mathrm{o}_{3/2}$ && 3s$^2$3p$^4$3d $^2$D$_{5/2}$  && {1.807e+11} \\  
&& 2-31 &&  175.263           &&   3s$^2$3p$^5$ $^2$P$^\mathrm{o}_{1/2}$ && 3s$^2$3p$^4$3d $^2$D$_{3/2}$  && {1.697e+11} \\  
Ref. lines &&1-28 &&  177.240           &&   3s$^2$3p$^5$ $^2$P$^\mathrm{o}_{3/2}$ && 3s$^2$3p$^4$3d $^2$P$_{3/2}$  && {1.466e+11} \\  
 && 1-27 &&  184.537           &&   3s$^2$3p$^5$ $^2$P$^\mathrm{o}_{3/2}$ && 3s$^2$3p$^4$3d $^2$S$_{1/2}$  && {1.249e+11} \\  
&& 1-7  &&  255.393           &&   3s$^2$3p$^5$ $^2$P$^\mathrm{o}_{3/2}$ && 3s$^2$3p$^4$3d $^4$D$_{1/2}$  && {3.453e+06} \\  
&& 1-6  &&  256.398           &&   3s$^2$3p$^5$ $^2$P$^\mathrm{o}_{3/2}$ && 3s$^2$3p$^4$3d $^4$D$_{3/2}$  &&  {5.767e+06} \\  
\hline
Temperature diagnostics && 1-3  &&  345.738           &&   3s$^2$3p$^5$ $^2$P$^\mathrm{o}_{3/2}$ && 3s3p$^6$ $^2$S$_{1/2}$   && {2.996e+09}    \\  
\hline        
Doublet && 5-20 && 1603.348             &&   3s$^2$3p$^4$3d $^4$D$_{5/2}$ &&  3s$^2$3p$^4$3d $^2$G$_{7/2}$ && {9.376e+00}\\  
&& 4-20 && 1603.206             &&   3s$^2$3p$^4$3d $^4$D$_{7/2}$ &&  3s$^2$3p$^4$3d $^2$G$_{7/2}$ && {1.937e+01}\\  
\hline
\end{tabular}\\
\end{table*}

From \eqns{eq:mixing} and (\ref{eq:MIT-fs}), the transition probability of MIT, $A_\mathrm{MIT}$, depends on both the magnetic field strength, B, and the energy separation between the $\mathrm{^4D_{7/2}}$ and $\mathrm{^4D_{5/2}}$ levels, $\Delta E$ = $E(\mathrm{^4D_{5/2}})$ - $E(\mathrm{^4D_{7/2}})$:
\begin{eqnarray}
\label{eq:amit}
{ A_\mathrm{{MIT}}} \propto \frac{B^2}{(\Delta E)^2}
\end{eqnarray}
\fig{fig:mit-m2} shows the ratio of rates for the MIT and M2 transitions as a function of magnetic field strength.
It is obvious that the sensitivity of the MIT line to the magnetic field strength is strongly dependent on the energy separation $\Delta E$ in the pseudo-degeneracy.
As one of the most critical parameters, the energy separation $\Delta E$ was estimated to be only a few cm$^{-1}$, which is beyond the limit of the accuracy of current theoretical calculations and is also far beyond the resolution power of any available EUV spectrometers.
There are several measurements of $\Delta E$ values from solar observations or laboratory measurements.
The first $\Delta E$ value was reported to be 5 cm$^{-1}$ from the analysis of the solar spectra observed by the Apollo Telescope Mount experiment S082B \citep{Bartoe1977} on board Skylab by \cite{Sandlin1979}.
Subsequently, observations from SERTS-89 \citep{Thomas1994} and SERTS-95 \citep{Brosius1998} predicted the same energy for the $\mathrm{^4D_{5/2}}$ and $\mathrm{^4D_{7/2}}$ levels due to the limited spectral resolution.
Inspired by the MIT project, \cite{Li2016} measured the $\Delta E$ value using an EBIT by measuring the line intensity ratio between the 257 \AA~blend and 256.398 \AA~line and comparing the measured value with a model using different values of splitting and inclusion of the MIT. 
They obtained a value of 3.5 cm$^{-1}$ with an upper limit of 7.8 cm$^{-1}$. 
A similar method has also been employed by \cite{Xu2022} and they determined $\Delta E$ = 8.6 $\pm$ 2.7 cm$^{-1}$, which is, however, much larger than that from \cite{Li2016}. The authors discussed the uncertainties from atomic data, spectrometer efficiency calibration and data statistics. Since this method is highly dependent on the accuracy of the collisional–radiative model (CRM), the atomic processes between different charge states, such as direct ionization, radiative recombination, and charge exchange that may happen under EBIT plasma conditions can also affect the populations between energy levels and, therefore, change the derived $\Delta E$ value significantly. However, these processes were not included in the CRM model used by \cite{Li2016} and \cite{Xu2022}. Hence all these considerations need to be carefully investigated in order to obtain an accurate value of $\Delta E$ from laboratory measurements in the future.

Another method to measure the $\Delta E$ value is by the analysis of the Fe~{\sc{x}} doublet at a longer wavelength region, which decay from the same upper level to the two lower quartet levels $\mathrm{^4D_{7/2,5/2}}$; 
the requirements on the spectral resolution are much less daunting by working at longer wavelengths. 
\cite{Judge2016} obtained a value of $\Delta E$ = 3.6 $\pm$ 2.9 cm$^{-1}$ by an analysis of S082B observations of UV spectra close to 1603.2 \AA, which decay from $\mathrm{3p^43d~^2G_{7/2}}$ to these two lower levels (see Table \ref{tab:lines}). 
More recently, \cite{Landi2020b} reported a new measurement of $\Delta E$ utilizing the 1603.2 \AA~spectra taken with the Solar Ultraviolet Measurement of Emitted Radiation \citep[SUMER,][]{SUMER} on board the Solar and Heliospheric Observatory, and obtained $\Delta E$ = 2.29 $\pm$ 0.5 cm$^{-1}$.
This new value agrees with and improves the measurements by \cite{Judge2016} and has been adopted for some recent studies for magnetic field measurements using the MIT method.
However, due to the critical importance of the $\Delta E$ value, more analysis of solar spectral observations as well as laboratory measurements are still needed to improve the accuracy of the $\Delta E$ value and therefore the magnetic field measurement.

\subsection{Transition data}

{Collisional and radiative processes dominate the atomic populations for typical electron densities of the active regions in the corona.}
Therefore, the accuracy of transition data and collision excitation rates involving higher energy levels are of great importance for the determination of line intensities/ratios used for the magnetic field strengths diagnostics.
The transition data in CHIANTI database (v10.0, \citep{chianti9,chianti10}) originate from \cite{DelZanna2012}, in which the atomic structure calculations were carried out using the autostructure program \citep{Badnell1997} which constructs target wavefunctions using radial wavefunctions calculated in a scaled Thomas-Fermi-Dirac statistical model potential with a set of scaling parameters.
More recently, \cite{Wang2020} carried out a large-scale ab initio MCDHF calculation for energy levels and radiative data for the lower n = 3 states in Cl-like ions including Fe~{\sc{x}}. \cite{Li_2021} further expanded the MCDHF calculation in Fe~{\sc{x}} to n = 4 levels.

Significant differences were found in the transition data between results from \cite{DelZanna2012} and \cite{Wang2020}; these differences can have significant effects on the level population and therefore the line intensities. 
For example, the measurements using an Fe~{\sc{x}} model that
combined the \cite{Wang2020} Einstein coefficients with the CHIANTI v9. collisional data increased the measured magnetic field strength by 20\%–30\%, compared to the results obtained utilizing the CHIANTI v9. data for both collisional and radiative data.
The radiative data from MCDHF calculations by \cite{Wang2020} and \cite{Li_2021} would be recommended for the spectral modeling of Fe~{\sc{x}} lines.
\subsection{Collisional data}
A number of studies of collisional strengths in Fe~{\sc{x}} have been presented in the literature, from mainly two theoretical methods, i.e., distorted-wave method and R-matrix method.
\cite{Mason1975,Malinovsky1980} and \cite{Bhatia1995} conducted the calculation of collision strength for Fe~{\sc{x}} using the University College London (UCL) distorted-wave code \citep{Eissner1998}.
\cite{Tayal2001} calculated the electron collision excitation strengths for transitions between the 49 lowest fine-structure levels using the semirelativistic R-matrix approach which took into account parts of the Breit-Pauli Hamiltonian.
Within the Iron Project, \cite{Pelan2001} performed a full Breit-Pauli R-matrix calculation for 180 levels arising from the five lowest n = 3 configurations.
Subsequently, \cite{Aggarwal2005} carried out a Dirac Atomic R-matrix Code (DARC, \cite{Grant1980}) calculation for the lowest 90 levels.
In addition, within the UK APAP network, \cite{DelZanna2004,DelZanna2012} performed detailed studies of collisional data in Fe~{\sc{x}}; their data have been adopted within CHIANTI database since 2005 \citep{Dere1997,Landi2006}.
The latest version of the CHIANTI database (v10.0, \citep{chianti9,chianti10}) used the collisional data from \cite{DelZanna2012}.
Recently, \cite{Li2022} did a new large-scale relativistic parallel DARC \citep{Ballance_DARC} calculation for the 
electron collisional excitation among 100 Fe~{\sc{x}}~levels. 
However, some discrepancies were found in the density and magnetic field measurements between the results obtained with different collisional data. 
Further efforts are still necessary on both theoretical and experimental sides to improve the accuracy of the atomic data, in order to provide a better estimation of the magnetic field using the MIT method.

\subsection{Laboratory measurement}
\begin{figure}
\centering
\includegraphics[width=\textwidth]{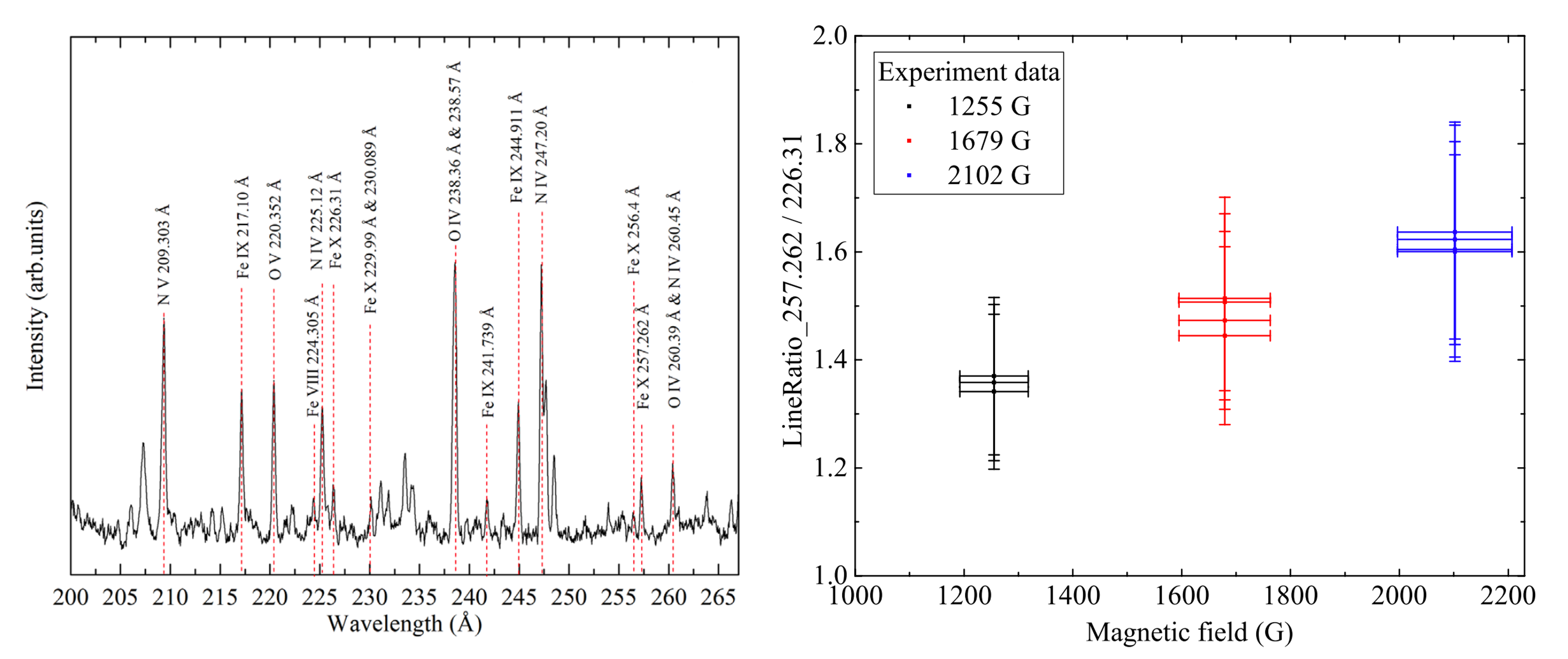}
\caption{Left: Spectra of Fe X taken at the Shanghai-HtscEBIT by using a grazing-incidence flat-field spectrometer. Right: The measured line ratios of 257/226 for Fe X ions at different magnetic fields. Images reproduced from \citet{Xu2022}.
}\label{fig:exp_EBIT}
\end{figure}

The effect of the MIT in Fe X at different magnetic fields has been verified experimentally in the laboratory by \citet{Xu2022}.
The measurements were performed at the Shanghai-EBIT laboratory by employing a grazing-incidence flat-field spectrometer installed on Shanghai-HtscEBIT.
The EUV spectra of Fe X in the wavelength range of 174–267 Å were collected at different magnetic field strengths and the representative spectrum for the longer-wave band 200–267 \AA~is shown in \fig{fig:exp_EBIT}.
The effective electron densities were obtained experimentally by measuring the widths of the electron beam and ion cloud with a grazing-incidence flat-field
spectrometer and visible spectrometer, respectively.
The measured electron density was further verified by the density-sensitive line ratios 174/175 and a good agreement was found between the two methods.
The magnetic field strength in Shanghai-HtscEBIT can be flexibly adjusted by setting the current of the coils and the spectra were recorded at three magnetic field strengths, i.e., 1255 G, 1679 G, and 2102 G. 
Then the line ratios between the 257 \AA~(MIT+M2+E1) and the reference line of 226.31 \AA~(257/226) were obtained at different magnetic field strengths and the results are shown in  \fig{fig:exp_EBIT}.
A noticeable trend can be found in the figure, where the magnetic field varies by a factor of roughly 1.6 and the line ratio 257/226 changes by 20\%, verifying the theoretical prediction based on the MIT technique. 
The sensitivity of the line ratios to the magnetic field reveal the potential capability of the MIT technique as a probe for diagnosing the plasma magnetic field.

From the figure, however, we can see that the line ratios 257/226 are associated with large uncertainties from Gaussian fitting and spectrometer efficiency calibration. Besides, the blending in the reference line 226.31 \AA~also introduced uncertainty in the measurement. Measurements using a higher resolution spectrometer is desirable in the future.
In addition, the magnetic fields in \cite{Xu2022} are much higher than that expected in the corona. Therefore, a more dedicated EBIT with lower magnetic field strengths might be needed.
The synchronous measurements of more line ratios between 257 \AA~and other reference lines, especially lines 174, 175, 177, and 184 \AA~ used in Hinode/EIS observations, are also highly desirable.

\section{Forward modeling of solar and stellar coronal magnetic field measurements}\label{sect:for_mod}

\begin{figure}
\centering
\includegraphics[width=\textwidth]{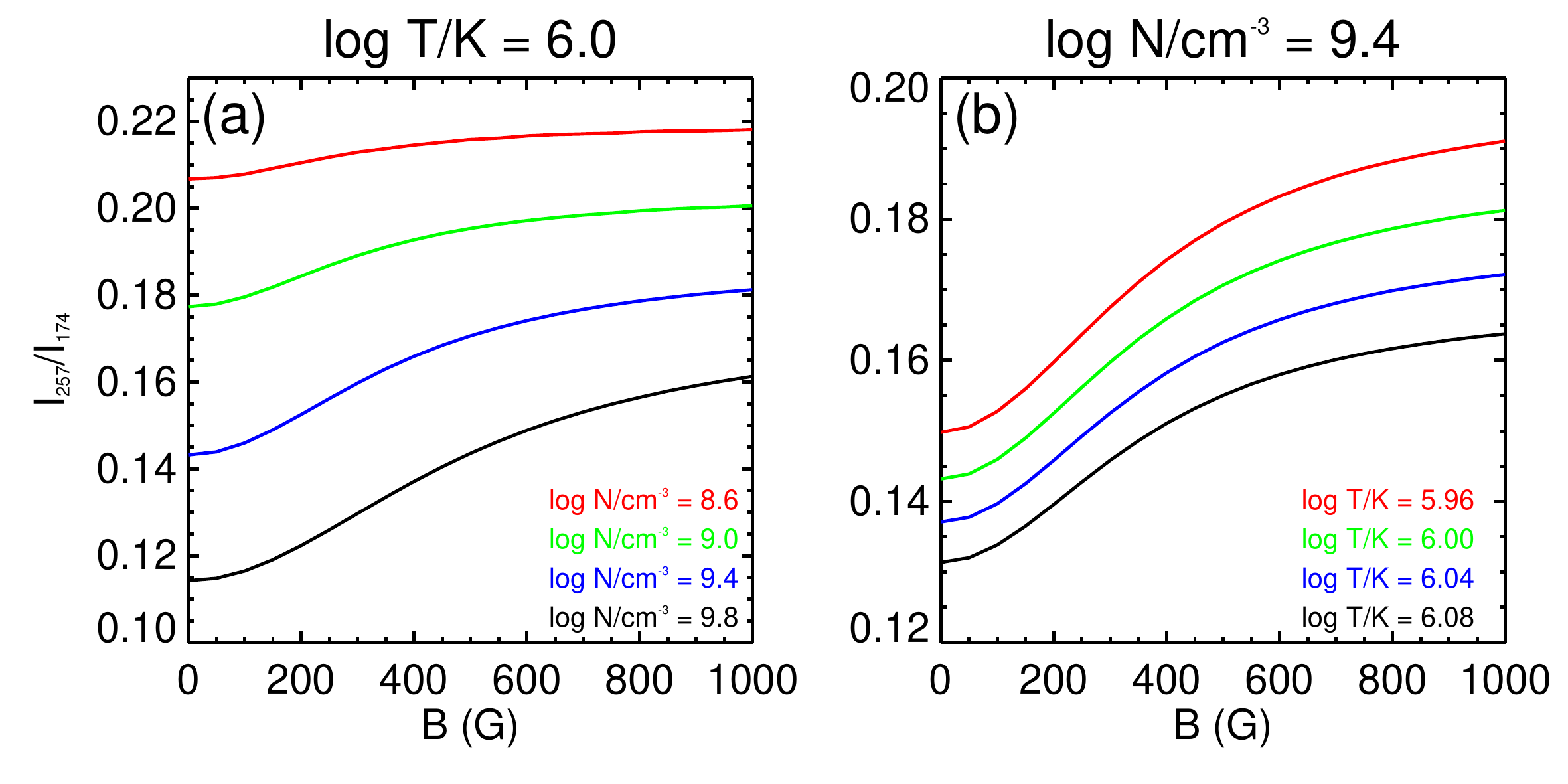}
\caption{Intensity ratio of the Fe~{\sc{x}} 257 and 174 {\AA} lines as a function of the magnetic field strength at different temperatures and densities. Image reproduced from \citet{Chen2021a}.
}\label{fig:lr_257_174}
\end{figure}

%
The MIT method is based on a simple concept: the intensity of the MIT emission and thereby the 257 \AA~(MIT+M2+E1) line intensity is directly affected by the local magnetic field strength;
therefore the intensity ratio of the 257 \AA~line and a reference Fe~{\sc{x}} line that is not sensitive to the magnetic field, 257/ref., can be used to measure the magnetic field strength.
The Fe~{\sc{x}} 174, 175, 177, 184, and 255 \AA~lines given in Table \ref{tab:lines} can be observed simultaneously along with the 257 \AA~line by Hinode/EIS, 
and therefore could be adopted as candidates of the reference lines \citep{Chen2021a}.
However, {it should be noticed} that the upper $^4$D$_{7/2}$ level is metastable, so any intensity ratios involving the 257 {\AA} line are also density-sensitive \citep{DelZanna2018}.
As an example, \fig{fig:lr_257_174} shows the intensity ratio of the Fe~{\sc{x}} 257/174 {\AA} line pair as a function of the magnetic field strength at different densities and temperatures.
Thus, the temperature and electron density must be determined before {the magnetic field inference via} the MIT method.
The density can be estimated by intensity ratios of density-sensitive line pairs, which are not significantly affected by magnetic field, such as the Fe~{\sc{x}} 174/175 {\AA} line pair.
The temperature of 10$^{6.0}$ K, at which the contribution functions of the Fe~{\sc{x}} lines peak, is often used as the formation temperature of the Fe~{\sc{x}} lines \citep[e.g.,][]{Si2020,Landi2020a,Brooks2021}.
More accurate temperature diagnostics can be achieved through the intensity ratio of the Fe~{\sc{x}} 184/345 {\AA} line pair \citep{Chen2021a}.

The suitability of the MIT method for solar and stellar coronal magnetic field strength measurements can be verified through forward modeling with 3D MHD models.
First, the emissions of the Fe~{\sc{x}} lines are synthesized from the models. Second, the density and temperature are determined from different line pairs.
Then, the magnetic field strengths can be derived based on the MIT method.
{The accuracy of the method can be evaluated by comparing the inferred values to those in the models.}
%
Such a procedure has been applied to models of a solar active region in \citet{Chen2021a} and global stellar coronae in \citet{Chen2021b} and \citet{Liu2022} to verify the feasibility of the MIT method for measuring the solar and stellar coronal magnetic field strengths.
\subsection{Solar coronal magnetic field measurements}

\begin{figure}
\centering
\includegraphics[width=12cm]{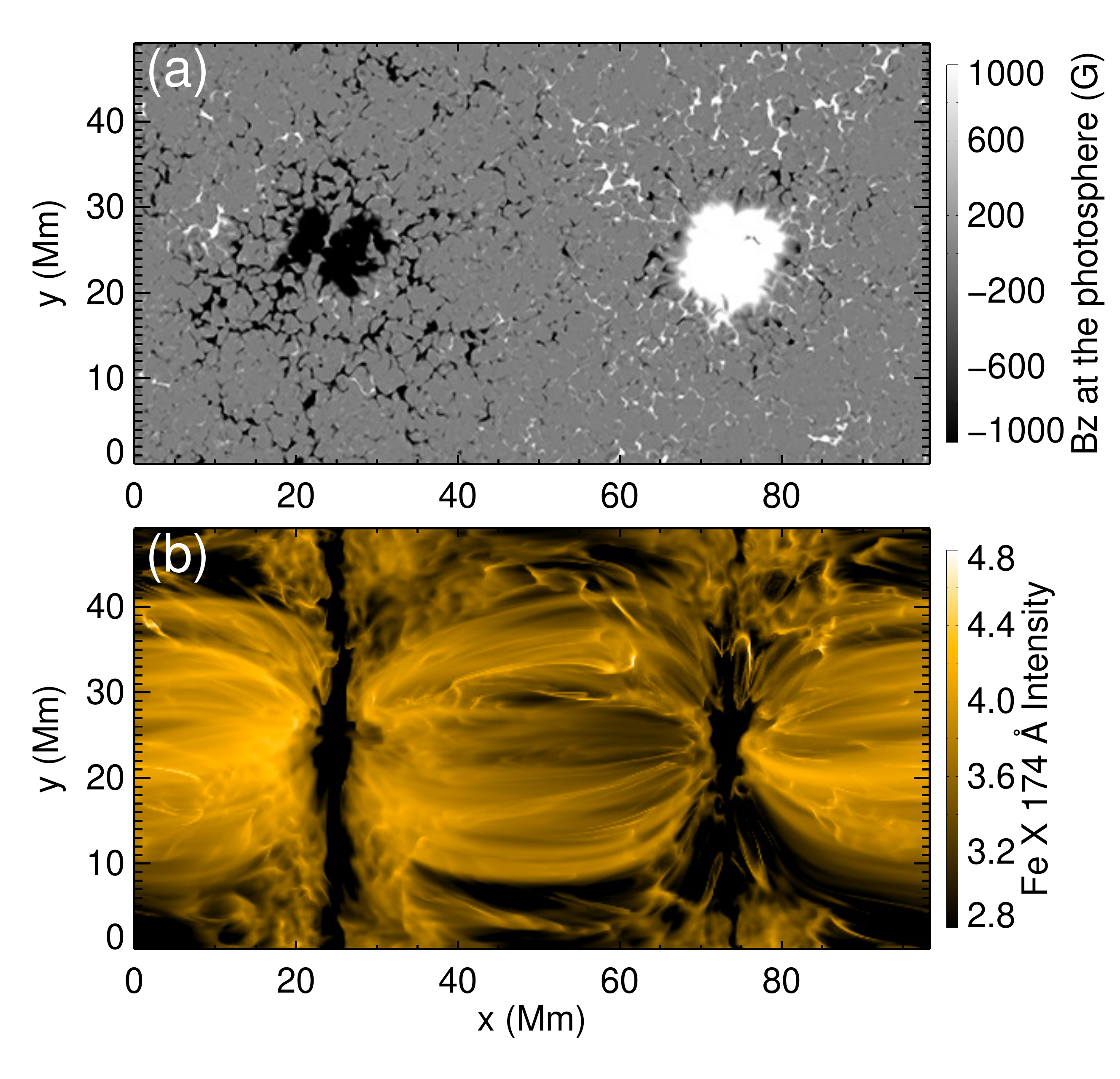}
\caption{(a) Vertical component of the photospheric magnetic field and (b) intensity map of the Fe~{\sc{x}} 174 {\AA} line obtained from a solar active region model. Image reproduced from \citet{Chen2021a}.
}\label{fig:ar}
\end{figure}

\citet{Chen2021a} took a 3D MHD model of a solar active region containing a bipolar sunspot pair. The coronal temperatures in the model are of the order of $\sim10^{5.9-6.2}$ K, which are typical values in the solar corona. The vertical component of the magnetic field in the photosphere and synthesized intensity map of the Fe~{\sc{x}} 174 {\AA} line when taking a line of sight (LOS) along the vertical direction are shown in \fig{fig:ar}. The coronal loop structures are clearly present in the synthesized coronal image.
%

\begin{figure}
\centering
\includegraphics[width=10cm]{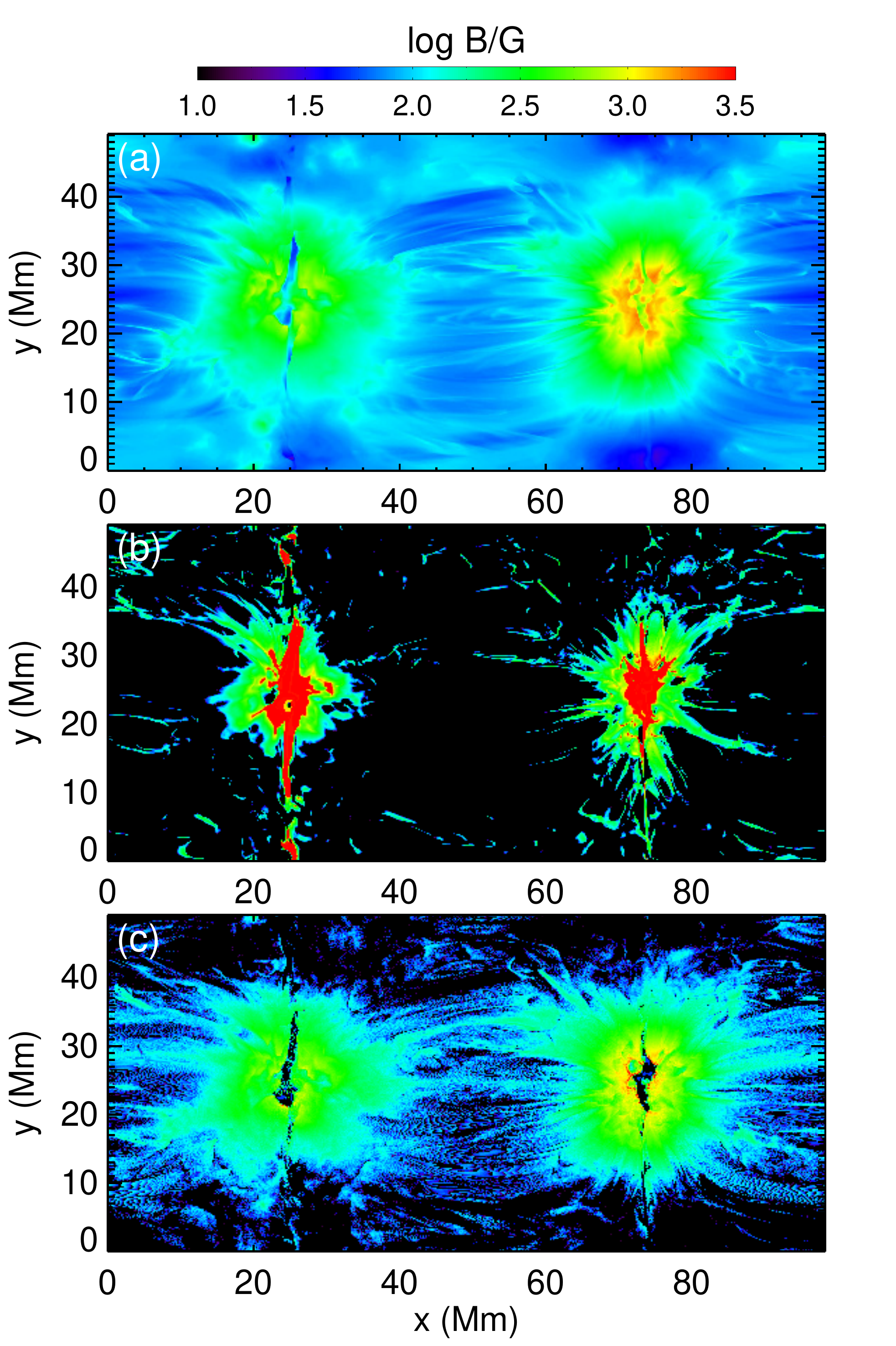}
\caption{(a) Coronal magnetic field strength in the model of a solar active region. (b) MIT-measured coronal magnetic field strength using intensity ratios of the Fe~{\sc{x}} 257/174 {\AA} line pair. The temperature is taken as 10$^{6.0}$ K when calculating electron density and magnetic field strength. (c) Similar to panel (b) but using Fe~{\sc{x}} 174/175 and 184/345 {\AA} line pairs for temperature and density diagnostics. Image reproduced from \citet{Chen2021a}.
}\label{fig:ar_ondisk}
\end{figure}

\begin{figure}
\centering
\includegraphics[width=\textwidth]{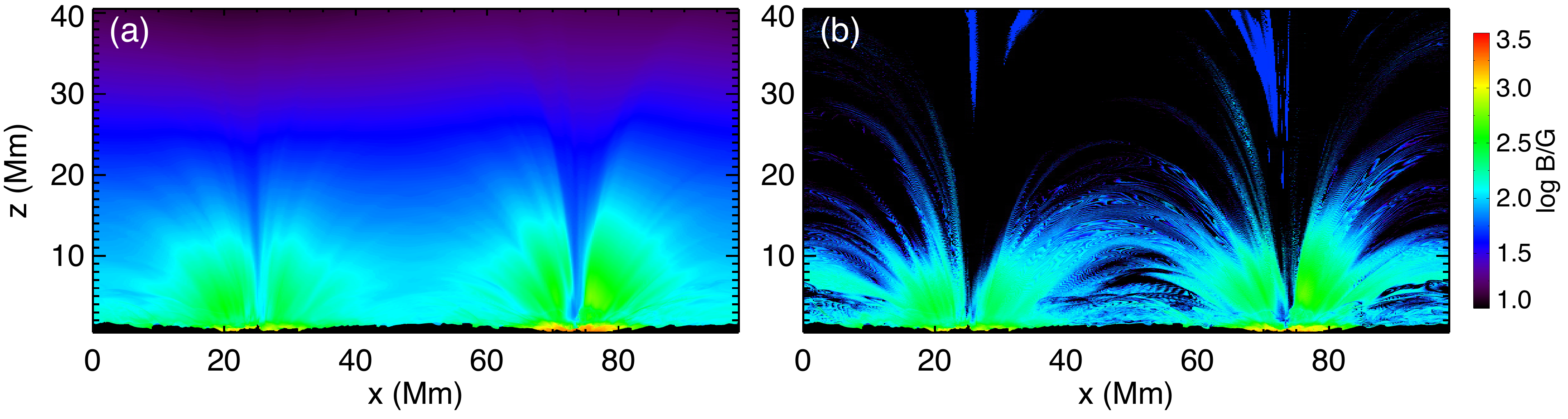}
\caption{Similar to \fig{fig:ar_ondisk} but for the mimicked off-limb observations. Image reproduced from \citet{Chen2021a}.
}\label{fig:ar_offlimb}
\end{figure}

In \citet{Chen2021a}, the authors employed two methods to evaluate the temperature and electron density.
The first one assumes a fixed temperature of 10$^{6.0}$ K at each pixel and then calculates electron density using the Fe~{\sc{x}} 174/175 {\AA} line ratio.
The second one is to derive the temperature and electron density simultaneously from the intensity ratios of 174/175 and 184/345 {\AA} line pairs based on a least-squares algorithm.
After the determination of temperature and density, the magnetic field strength can be derived from the 257/ref. ratios.
Note that since the Fe~{\sc{x}} lines are optically thin, the line intensity is the integrated emissivity along the LOS. Thus, the coronal magnetic field strength in the model can be defined as the emission-weighted averaged field strength ($B_0$).
The comparisons between the coronal magnetic field strengths in the model and MIT-measured values using 257/174 {\AA} line ratio from the two methods of temperature measurements are shown in \fig{fig:ar_ondisk}.
We can see that the assumption of a fixed temperature of 10$^{6.0}$ K results in the failure of the magnetic field measurement in regions away from the sunspots and introduces significant uncertainties in the magnetic field strength measurements using the MIT method.
On the contrary, the least-squares algorithm provides a more accurate estimation of the magnetic field strength, for not only the regions around the footpoints but also in higher parts of the coronal loops.
Therefore, simultaneous temperature and density determination is important for coronal magnetic field measurements using the MIT method.
Furthermore, they applied the same method to the mimicked off-limb observation of an active region and the results are shown in \fig{fig:ar_offlimb}.
From the figure we can see that the MIT method can provide a reasonably accurate estimation of the magnetic field strength in the coronal loop structures. 
For the regions with magnetic field strength stronger than 150 G, the differences between the MIT-measured field strengths and the values in the model are mostly below $\sim$30\%.
Overall, the MIT method using 257/ref. ratios can provide reasonably good estimations of coronal magnetic field strength in both on-disk and off-limb observations.

It is worth noting that \citet{Chen2021a} did not consider the magnetic field strength when diagnosing the temperature and density.
However, the intensity ratios of 174/175 and 184/345 {\AA} line pairs also slightly change with magnetic field strength.
The ignorance of the magnetic field strength results in slight underestimation of temperature and density in the regions with field strength stronger than $\sim$600 G, and thereby underestimates the magnetic field strength within the sunspots.
Recently, \citet{Martinez2022} proposed a new approach to derive the density, temperature, and magnetic field simultaneously.
The new method employs the same spectral lines used by \cite{Chen2021a}, i.e., 174, 175, 184, 257, 345 {\AA} lines listed in Table \ref{tab:lines}.
They took account in the contribution functions of these Fe~{\sc{x}} lines as functions of temperature, density, and magnetic field strength and calculated the differential emission measure (DEM) for each temperature, density, and magnetic field strength based on the inversion method developed by \citet{DEM_Cheung}.
Through comparison between the values derived from the MIT technique and the emissivity-weighted magnetic field strength, they found this method can provide reasonably accurate coronal magnetic field strength for the regions without significant bound-free absorption (the effects of absorption will be further discussed in \sect{sect:limitations}).
Furthermore, their technique allows the identification of different coronal structures with different magnetic field strengths along the LOS.

\subsection{Measurements of the magnetic field at the bases of stellar coronae}

\begin{figure}
\centering
\includegraphics[width=\textwidth]{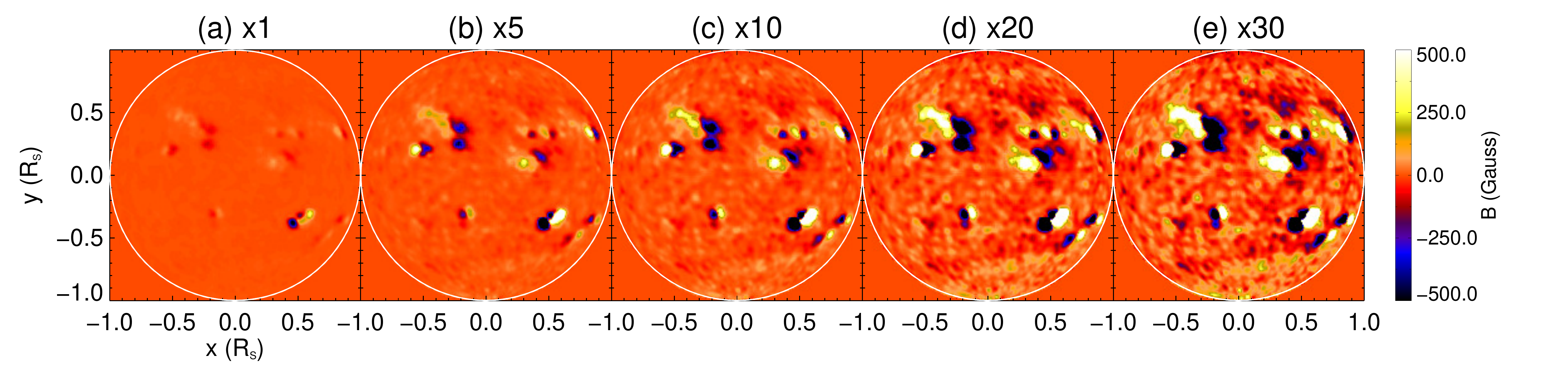}
\caption{Radial component of the magnetic field at the surface in different stellar models. The white circles indicate the limb of the photosphere. Image reproduced from \citet{Chen2021b}.
}\label{fig:stellar1}
\end{figure}

\begin{figure}
\centering
\includegraphics[width=\textwidth]{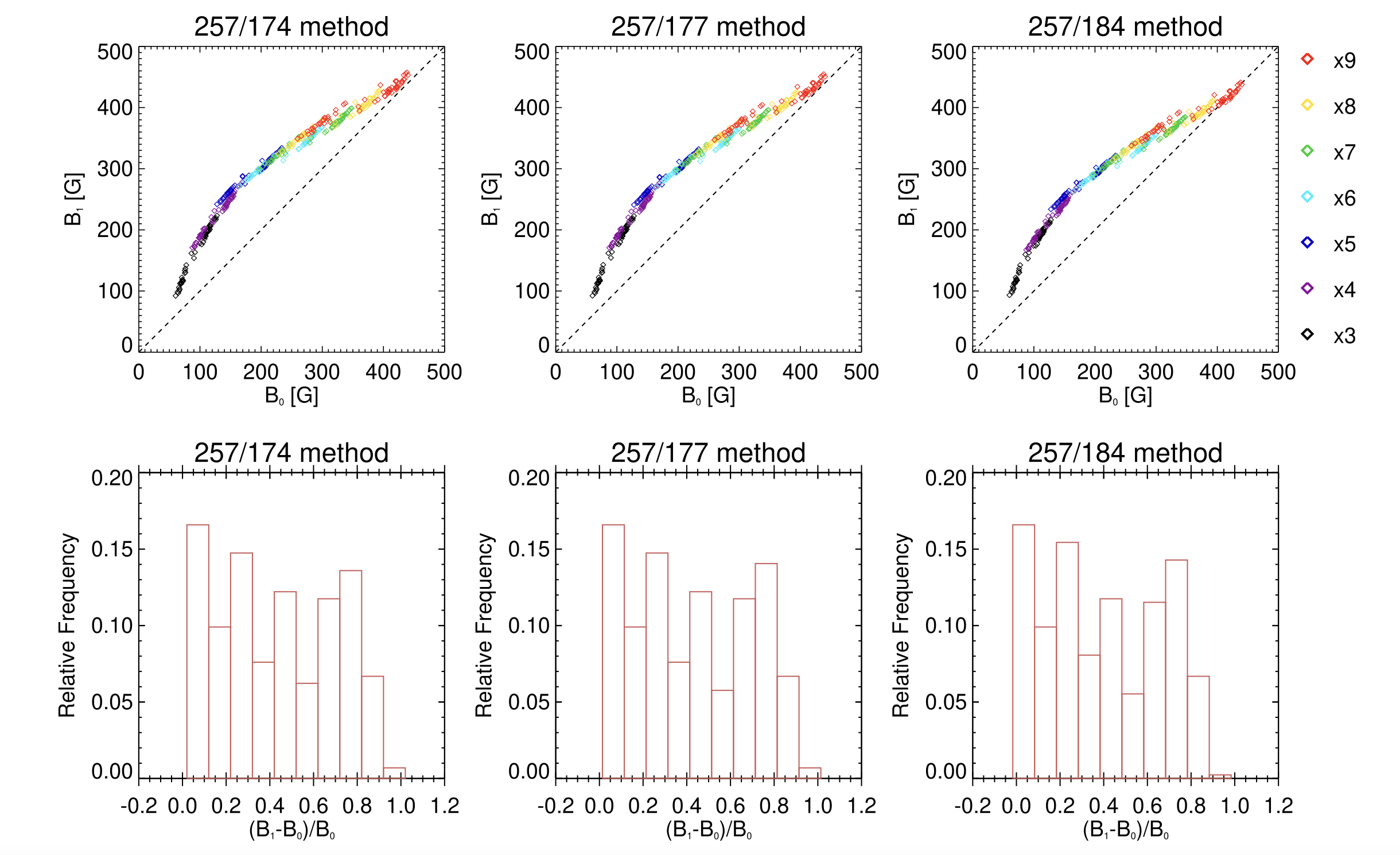}
\caption{Comparison between MIT-measured field strength (B$_1$) and the values in the models (B$_0$).
Upper panels are the scatter plots of B$_0$ and B$_1$ derived from different line ratios. The different colors represent results from different models, and different data points with the same color correspond to different LOS directions.
Lower panels are the histograms of differences between B$_0$ and B$_1$. Image reproduced from \citet{Liu2022}.
}\label{fig:stellar2}
\end{figure}

To investigate the possibility of extending the MIT method to measurements of coronal magnetic field strength in other late-type stars, \citet{Chen2021b} performed forward modeling with a series of steady-state global MHD models produced by \citet{Jin2020}. 
These stellar models are constructed by multiplying the surface magnetic flux density of the real solar observations taken during the ascending phase of solar cycle 24 by factors of 1, 5, 10, 20, and 30, respectively.
The radial component of the magnetic field at the stellar surfaces is presented in \fig{fig:stellar1}. 
In \cite{Chen2021b}, the synthesized emissions of the Fe~{\sc{x}} lines are {integrated over the whole star (backside excluded)} as the stars are spatially unresolved. For each model, they estimated the electron density from the 174/175 {\AA} line ratio and then calculated the magnetic field strength using intensity ratios of the 257 {\AA} line and other reference lines.
By comparing the magnetic field strengths in the models with MIT-measured values,
\citet{Chen2021b} found that the differences between the measured magnetic field strengths and the values in the model are mostly smaller than 50\% for the stars with a mean surface magnetic flux density more than 20 times higher than that of the Sun, indicating that the MIT diagnostic technique can provide reasonably accurate magnetic field strength measurements.

In \citet{Chen2021b}, the input photospheric magnetograms of the MHD models were scaled from a solar photospheric magnetogram taken during a relatively inactive phase of the solar cycle, which may not be appropriate for investigations of magnetic fields on active stars or solar-type stars at the peak of their long-term activity cycles \citep{Wilson1978}.
More recently, \citet{Liu2022} constructed a series of stellar MHD models, in which the photospheric magnetograms are scaled from a solar synchronous magnetogram taken during the solar maximum.
They applied the MIT technique in \citet{Chen2021b} and the results are shown in \fig{fig:stellar2}. 
The MIT method can be used to diagnose the magnetic field strength of stars with a magnetic flux density of at least 3 times higher than that of the Sun at the solar maximum, and the differences between the MIT-measured field strengths and the values in the models are less than a factor of 2.
It is worth noting that the average coronal temperatures in the stellar models of \citet{Chen2021b} and \citet{Liu2022} reach up to 10$^{6.5}$ K. Considering that the contribution functions of the Fe~{\sc{x}} lines peak at 10$^{6.0}$ K, these Fe~{\sc{x}} emissions are mainly from the base of stellar coronae.
Thus, according to the forward modeling of \citet{Chen2021a,Chen2021b} and \citet{Liu2022}, the MIT method has the potential to measure the magnetic field strength at the coronal bases of some nearby stars with a mean surface magnetic flux density a few times higher than that of the Sun.

\section{Application to \textit{Hinode}/EIS spectroscopic observations}\label{sect:EIS}

%
The Fe~{\sc{x}} 257, 174, 175, 177, 184, 255 {\AA} lines have been routinely observed by \textit{Hinode}/EIS., which makes it possible to measure the coronal magnetic field strength by applying the MIT method to real solar observations.
In the last three years, the MIT diagnostic method has been applied to various EIS observations for coronal magnetic field strength measurements in solar active regions \citep{Si2020,Landi2020a,Landi2021,Brooks2021,Brooks2021b}. Though being subject to large uncertainties, these attempts are still very valuable.

\subsection{Direct line ratio technique}

\begin{figure}
\centering
\includegraphics[width=10cm]{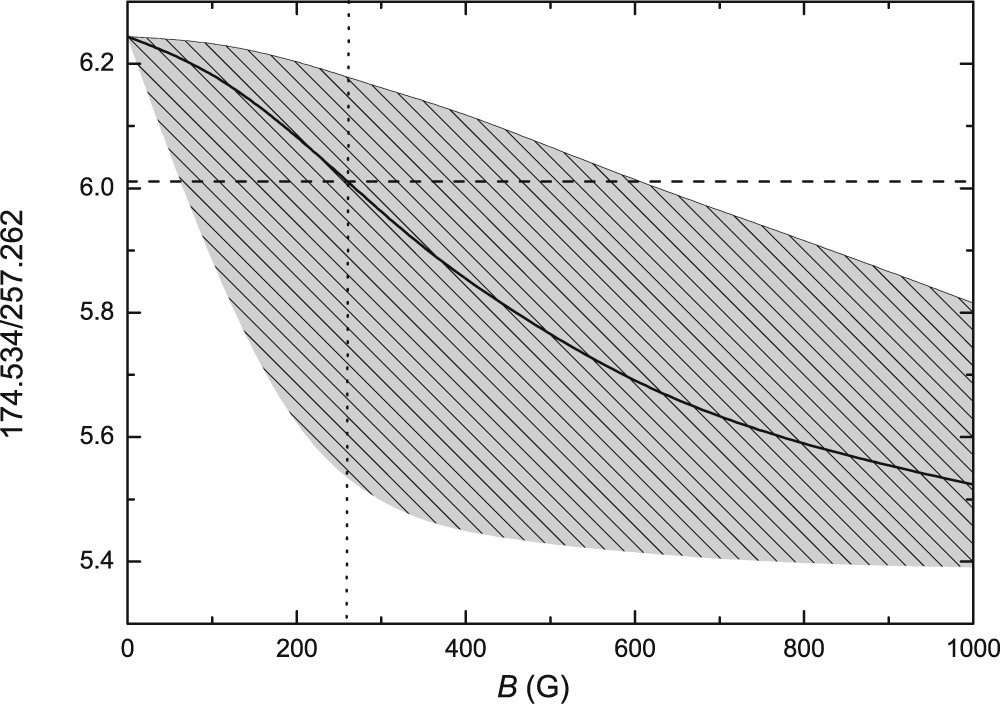}
\caption{Intensity ratio of the 174/257 {\AA} line pair as a function of the magnetic field. The gray area represents the uncertainty caused by the uncertainty in the value of $\Delta E$. The horizontal dashed line indicates the measured intensity ratio of the 174/257 {\AA} line pair in an active region. Image reproduced from \citet{Si2020}.
}\label{fig:si_MIT}
\end{figure}

The first application of the MIT method to determine the magnetic field strength in an active region from EIS observations was presented by \cite{Si2020}.
The intensities of the Fe~{\sc{x}} lines were taken from an active region, as given by \citet{Brown2008}.
They estimated the electron density from the Fe~{\sc{x}} 174/175 {\AA} line pair and then obtained a coronal magnetic field strength of 270 G by comparing intensity ratios of the 257 {\AA} line and a reference line (174 or 175 {\AA} line) between theoretical predictions and EIS observations (as seen in \fig{fig:si_MIT}).
The significant uncertainty in their magnetic field measurements
mainly comes from the uncertainty in $\Delta E$ value, which was taken as
3.6$\pm$2.7 cm$^{-1}$ from \citet{Judge2016}.
In the direct line ratio technique, the blend Fe~{\sc{x}} 257 {\AA} line emission is dominated by the E1 transition, especially when the magnetic field strengths is weak; therefore the sensitivity of this technique is limited and it poses severe requirement on the accuracy of atomic data and intensity
calibration, when compared to the weak- and strong-field techniques (see Section \ref{sect:WF&SF}).

\subsection{Weak-field and strong-field techniques}\label{sect:WF&SF}

\begin{figure}
\centering
\includegraphics[width=\textwidth]{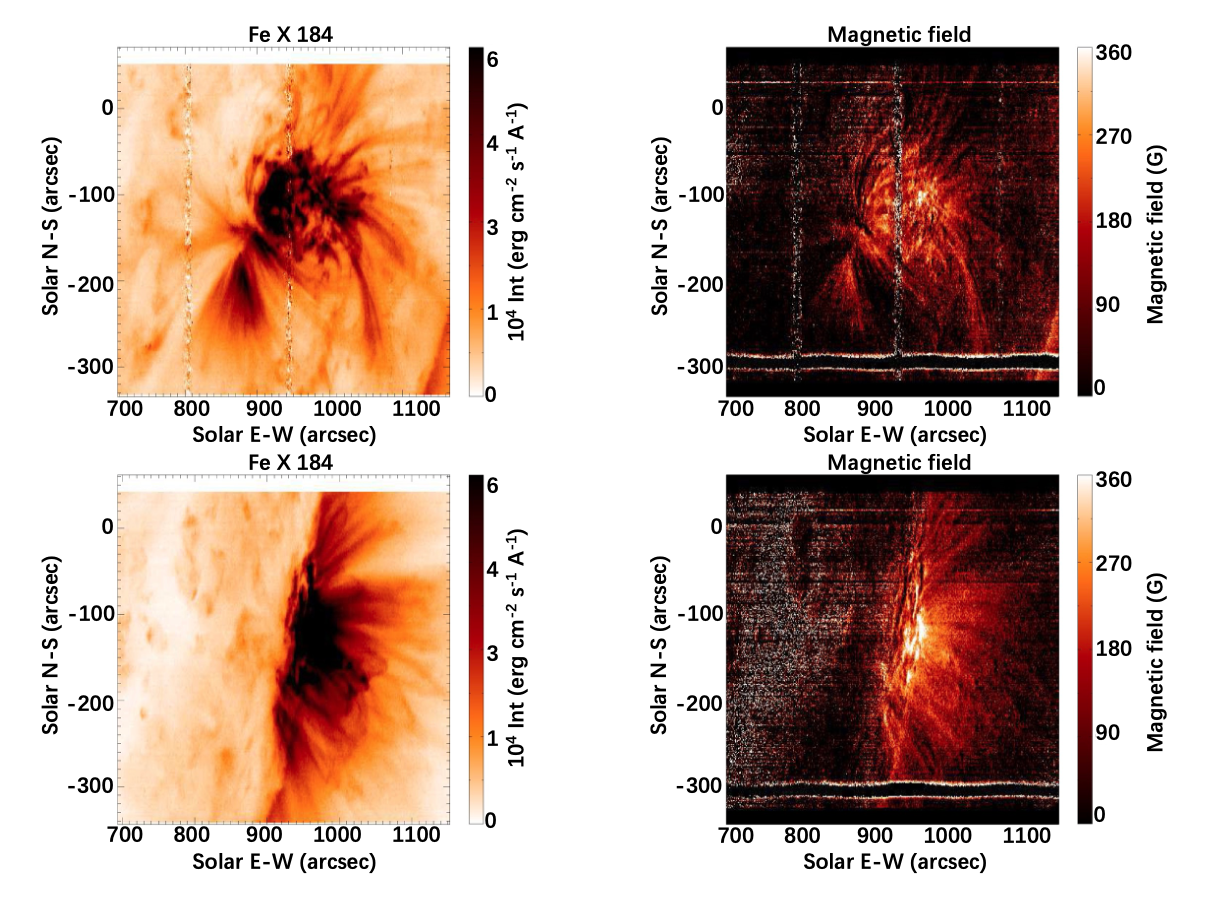}
\caption{Left: Intensity maps of the Fe~{\sc{x}} 184 {\AA} line. Right: Magnetic field maps derived from the weak field technique. Image reproduced from \citet{Landi2020a}.
}\label{fig:landi_MIT}
\end{figure}

%
\citet{Landi2020a} proposed two techniques, namely weak-field and strong-field techniques, that can be applied to weak- and strong-field situations, respectively.
For the weak-field technique, it is assumed that the MIT transition is weaker than the M2 transition and does not affect the level population of $\mathrm{^4D_{7/2}}$, and hence the M2 intensity.
This allows one to separate the MIT contribution from the
blend (E1+M2+MIT) intensity and derive the magnetic field strength by comparing the MIT/M2 branching ratio to theoretical prediction \citep{Landi2020a}:
\begin{eqnarray}
\label{eq:weak}
\frac{I(\rm MIT)}{I( \rm M2)} = \frac{I(257)}{I(\rm Ref.)} \cdot R(\frac{\rm Ref.}{\rm M2}) - R(\frac{\rm E1+M2}{\rm M2})
\end{eqnarray}
where $I(\rm 257)$ and $I(\rm Ref.)$ are the measured intensities of the MIT-blend 257 \AA~line and a reference Fe~{\sc{x}} line, respectively.
$R(\frac{\rm Ref.}{\rm M2})$ and $R(\frac{\rm E1+M2}{\rm M2})$ are theoretically predicted intensity ratios calculated using the CHIANTI database \citep{chianti10,Dere1997}, in which the MIT is not taken into account.
The weak-field technique is more sensitive to the field strength than the direct line ratio method (see \figs{fig:mit-m2} and \ref{fig:lr_257_174}), but is only adequate under the presence of weak field strength, e.g., less than 150-200 G as suggested in \cite{Landi2020a}.
As the magnetic field strength increases, the MIT transition rate will dominate over the M2 rate and affect the level population of $\mathrm{^4D_{7/2}}$;
therefore, the weak-field assumption is not valid any more.

\citet{Landi2020a} first applied the weak-field technique to both on-disk and off-limb EIS observations of some solar active regions.
For the density determinations, the best choice is using the Fe~{\sc{x}} 174/175 {\AA} line ratio.
However, these two lines are not always included in EIS observations, and the signal-to-noise ratios for these two lines are often very low because they are observed at the edge of the short-wavelength detector.
As the formation temperatures of Fe~{\sc{x}} and Fe~{\sc{xi}} are close,
\citet{Landi2020a} suggested that the Fe~{\sc{xi}} 182.17/(188.22+188.30) {\AA} line ratio can also be used for density diagnostics.
After deriving the electron density, the coronal magnetic field strength is estimated according to \eqn{eq:weak}.
Two examples of the results from \citet{Landi2020a} are shown in \fig{fig:landi_MIT}, which demonstrates that the weak-field technique can provide 2D magnetic field maps of active regions in both on-disk and off-limb observations.
Moreover, \cite{Landi2021} measured the evolution of the magnetic field before, during, and after a C2.0 flare using the weak-field technique and found that the flare is associated with a large magnetic field enhancement of $\sim$500 G.

In early studies, a constant temperature of 10$^{6}$ K is always assumed when applying the MIT method to EIS observations \citep[e.g.,][]{Si2020,Landi2020a,Landi2021}.
\citep{Chen2021a} has demonstrated that accurate temperature determination is of critical importance for coronal magnetic field measurements using the MIT method.
Recently, \citet{Brooks2021b} estimated the temperature from the Gaussian emission measure analysis, and then used this temperature for the calculation and obtained magnetic field strengths of 60-150 G in coronal loops.

\begin{figure}
\centering
\includegraphics[width=\textwidth]{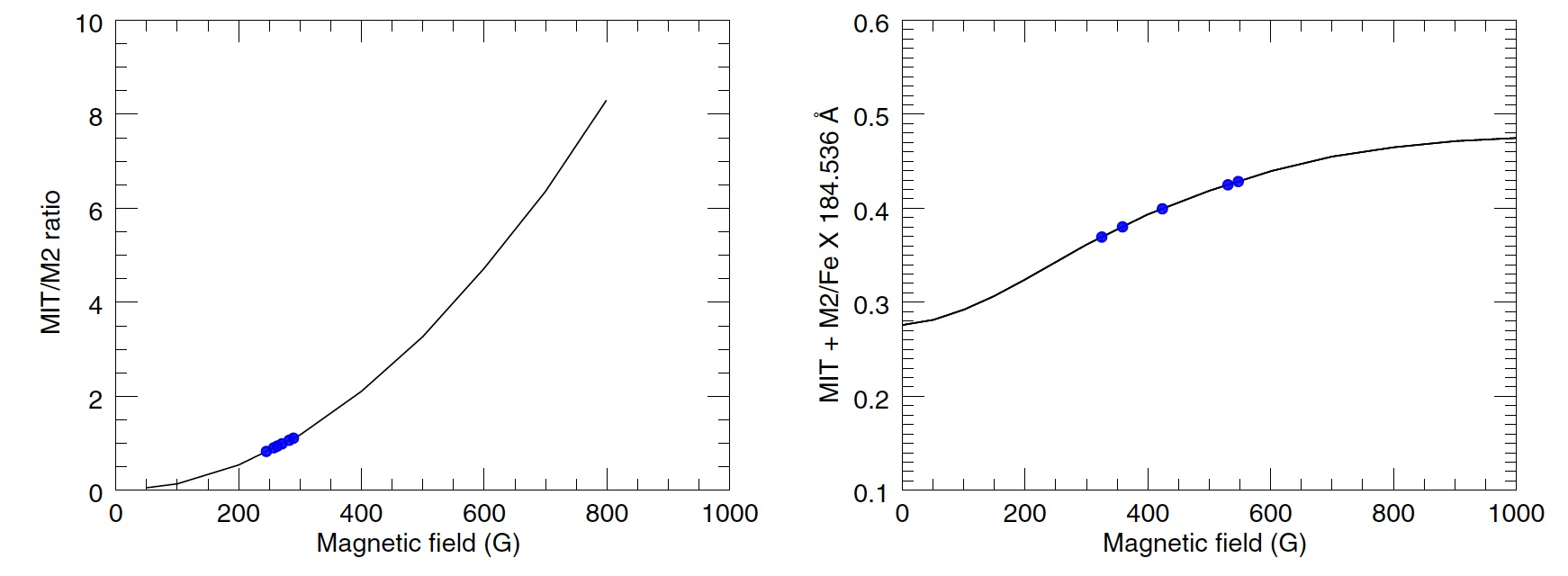}
\caption{Left: MIT/M2 intensity ratio as a function of magnetic field. Right: I(M2+MIT)/I(ref) ratio as a function of magnetic field. The blue dots in the left and right panels correspond to the magnetic field strength measured using \eqns{eq:weak} and \ref{eq:strong}, respectively. Image reproduced from \citet{Brooks2021}.
}\label{fig:wso}
\end{figure}

The strong-field technique considers the influence of MIT on the population of $\mathrm{^4D_{7/2}}$ but neglects the influence on the reference lines;
so that the 257.259 \AA~E1 intensity can be isolated from the blend 257 \AA~line intensity. 
Thus, the magnetic field strength can be obtained by calculating the I$\mathrm{{(MIT+M2)}}$/I$\mathrm{{(ref)}}$ ratio with CHIANTI database and comparing with observations \citep{Landi2020a}:
\begin{eqnarray}
\label{eq:strong}
\frac{I(\rm M2+MIT)}{I( \rm Ref.)} = \frac{I(257)}{I(\rm Ref.)} - R(\frac{\rm E1}{\rm Ref.})
\end{eqnarray}
where $R(\frac{\rm E1}{\rm Ref.})$ is the intensity ratio between the E1 component and the reference line predicted from CHIANTI. 
\cite{Brooks2021} calculated the coronal magnetic field strength of the active region 11944 using both the weak-field and strong-field techniques, and the results in the source region of solar energetic particles (SEPs) obtained from the different methods are shown in \fig{fig:wso}.
The coronal magnetic field strength in the confined source region of the SEPs is several hundred Gauss, and the plasma continually released from the confined magnetic field structure and accelerated as SEPs during flares.
Although the magnetic field strengths calculated from the strong-field technique are higher than the values derived from the weak-field technique, 
the results of the two methods are consistent considering the uncertainties from both the observations and the diagnostic methods themselves.

\section{Uncertainties, limitations, and future prospects} \label{sect:limitations}

Despite the great potential for magnetic field diagnostics, the application of the MIT method is currently subject to several uncertainties and limitations. The uncertainties in the atomic data and solar/stellar observations may affect the accuracy of coronal magnetic field strength measurements using the MIT method.

The precise measurement of $\Delta E$ is very challenging (see \sect{sect:delta_E}), and the most recent measurement gives a value of 2.29$\pm$0.5 cm$^{-1}$ \citep{Landi2020b}. 
This value has been used in forward modeling of the MIT method \citep[e.g.,][]{Chen2021a,Chen2021b,Liu2022} and coronal magnetic field measurements based on EIS observations \citep[e.g.,][]{Landi2020a,Brooks2021,Brooks2021b}.
\citet{Chen2021a} investigated the effect of the uncertainty in $\Delta E$ on the measurements of coronal magnetic field and found that the choice of $\Delta E$ values in the range of 1.79--2.79 cm$^{-1}$ does not significantly change the suitability of the MIT method but could lead to slight systematic deviations from the coronal magnetic field in the model.

\begin{figure}
\centering
\includegraphics[width=12.5cm]{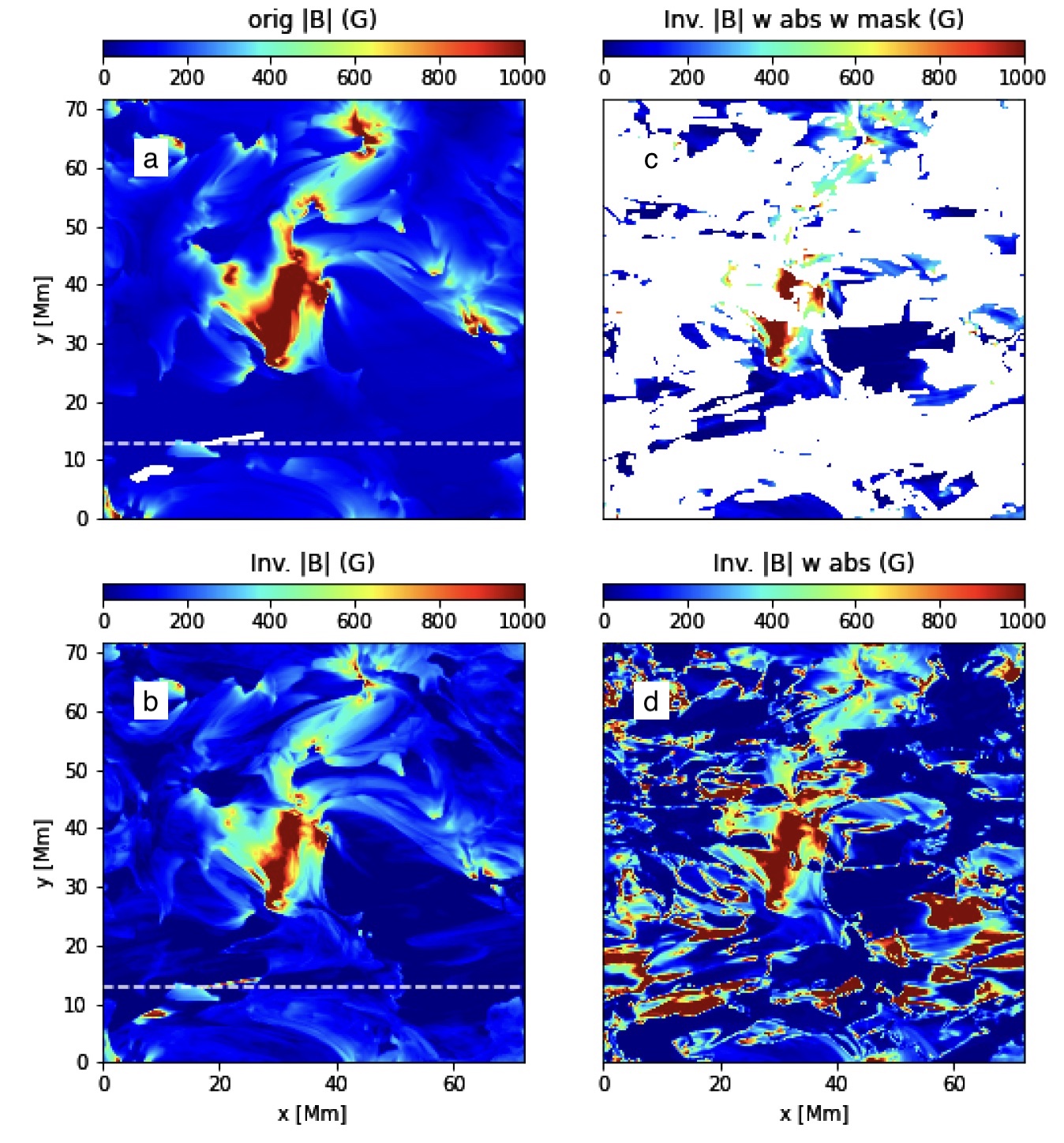}
\caption{Effects of absorption from cool plasma at coronal heights.
(a) The coronal magnetic field in the model.
(b) The MIT-measured magnetic field strengths without considering absorption.
(d) Similar to panel (b) but including bound-free absorption from cool plasma.
(c) Similar to panel (d) but masking out the area with strong absorption.
Image reproduced from \citet{Martinez2022}.
}\label{fig:absorption}
\end{figure}

Cool plasma may occur at the coronal heights \citep[e.g.,][]{Berger1999}, which can lead to absorption of emissions from the Fe~{\sc{x}} lines, especially in filaments \citep{Anzer2005}, moss regions \citep{DePontieu2009}, and regions with strong emerging magnetic flux \citep{Hansteen2019}.
\citet{Martinez2022} took into account bound-free absorption from neutral hydrogen, helium, and singly ionized helium when synthesizing the emissions of the Fe~{\sc{x}} lines from the MHD models.
They found that, for the model with strong flux emergence, some cool plasma with photospheric temperature is carried into coronal heights; the absorption of the Fe~{\sc{x}} emission from cool plasma significantly affects the accuracy of density and magnetic field diagnostics (as shown in \fig{fig:absorption}).
While for the regions without absorption, the MIT method can still provide reasonable estimations of the coronal magnetic field strength.

\begin{figure}
\centering
\includegraphics[width=\textwidth]{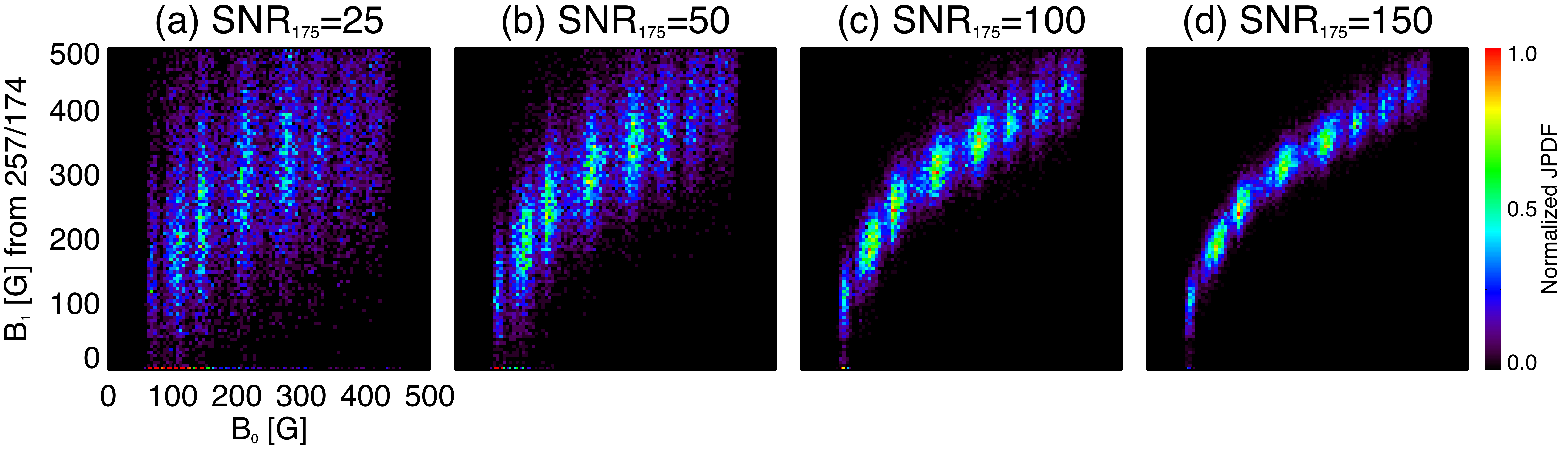}
\caption{Joint probability density functions of B$_0$ and B$_1$ calculated from 257/174 {\AA} line ratio. Panels (a--d) are the results under signal-to-noise ratios of 25, 50, 100, and 150, respectively.
Image reproduced from \citet{Liu2022}.
}\label{fig:stellar3}
\end{figure}

Uncertainties from spectral observations of the Fe~{\sc{x}} lines, such as measurement errors caused by photon counting and radiometric calibration, will inevitably impact the accuracy of the coronal magnetic field measurements.
\citet{Liu2022} investigated the uncertainty of magnetic field measurements from photon counting error and their results are shown in \fig{fig:stellar3}.
It is evident that a signal-to-noise ratio of 25 for the Fe~{\sc{x}} 175 {\AA} line, which is the weakest among the Fe x lines used in the diagnostics, is not sufficient to provide accurate coronal magnetic field strength measurements from the MIT technique.
For signal-to-noise ratios of 50 or higher, the uncertainties of MIT-measured values are 20-30\%. Thus, they proposed that a signal-to-noise ratio of at least 50 for the 175 {\AA} line is needed to achieve effective magnetic field measurements of stellar coronae using the MIT method.
An accurate relative intensity calibration among different Fe~{\sc{x}} lines in EIS observations is very important for precise measurements of magnetic field strengths.
However, the Fe~{\sc{x}} 257 {\AA} line is observed by the long-wavelength detector of EIS, while the adopted reference lines, such as the 174, 175, and 184 {\AA} lines, are observed by the short-wavelength detector.
This makes the accurate radiometric calibration rather challenging and could cause significant uncertainties in magnetic field measurements.
\cite{Landi2020a} employed two independent calibration methods developed by \citet{HPW} and \citet{GDZ}, respectively, and found large discrepancies between the two derived magnetic field strengths. 
They concluded that the intensity calibration uncertainty is at the moment the greatest obstacle to the application of the MIT method to EIS observations.

To obtain accurate density and temperature diagnostics, the best choice is to use at least the Fe~{\sc{x}} 174, 175, 184, 257, and 345 {\AA} lines \citep{Chen2021a,Martinez2022}.
However, the 345 {\AA} line is not covered by EIS, which makes it difficult for temperature diagnostics.
An alternative method is to derive the temperatures from the temperature-sensitive Fe {\sc{ix}} 171.07/188.49 {\AA} and Fe~{\sc{xi}} 188.22/257.55 {\AA} line pairs, respectively, and then take their average as the formation temperature of the Fe~{\sc{x}} lines.
Another approach to obtain the temperature information is based on the DEM analysis.
Nevertheless, \citet{Chen2022} performed forward modeling of both approaches, and found that the correlation between the MIT-measured magnetic field strength and the field strength in the model is much worse in both methods, when compared to the situation of temperature diagnostics from the Fe~{\sc{x}} lines. 
In other words, we still need to figure out a better way to estimate the formation temperature of the Fe~{\sc{x}} lines from EIS observations.
Besides, the Fe~{\sc{x}} 174 and 175 {\AA} lines are at the edge of the short-wavelength detector, meaning that the signal-to-noise ratios of the two lines are normally low in EIS observations and thus density diagnostics from these lines may be subject to large uncertainty.
As an alternative, the Fe~{\sc{xi}} 182/188(blend) {\AA} line pair has been used for density estimations \citep{Landi2020a,Brooks2021b}.
However, based on the forward modeling by \cite{Liu2022}, the Fe~{\sc{xi}} line pair results in a systematic underestimation of both the electron density and magnetic field strength.

Furthermore, \citet{Chen2022} performed forward modeling of the weak-field technique using the same algorithms employed by \cite{Chen2021a}: the first method is assuming a uniform temperature of 10$^{6.0}$ K and estimating the density using the Fe~{\sc{x}} 174/175 {\AA} line ratio; the second method is determining the temperature and density simultaneously from the 174/175 and 184/345 {\AA} line pairs based on a least-squares technique.
Compared to the results derived from the direct line ratio technique in \cite{Chen2021a}, the weak-field technique does not improve the suitability of the MIT method for the same given temperature and density maps.
And they found that the weak-field technique significantly underestimates the coronal magnetic field in the model.

In the end, the MIT diagnostic method can only provide the information about the magnetic field strength but not its direction. It is highly desirable to combine different magnetic field imaging techniques such as coronal seismology \cite[e.g.,][]{Yang2020a,Yang2020b} and spectropolarimetry \cite{Lin2004} to achieve a better and complete understanding of coronal magnetism.

\section{Summary}
Routine and accurate coronal magnetic field measurement is of great importance for our understanding of various physical processes in solar and stellar coronae and, in particular, helps us interpret up-to-date existing and upcoming observations obtained by recent space missions such as Solar Orbiter \citep{2020A&A...642A...1M} and Advanced Space-based Solar Observatory \citep[ASO-S, ][]{ASOS}.
Recently, {the MIT technique has attracted great attention due to its diagnostic potential for magnetic fields in the solar and stellar coronae.}
In this review, we have introduced the quantum theory underlying the MIT method, and summarized the calculations of the important atomic parameters critical for accurate magnetic field measurements.
Numerous efforts have been taken to improve the accuracy of the atomic data, such as the radiation transition rates, the collisional rates, and the energy separation between $\mathrm{^4D_{5/2}}$ and $\mathrm{^4D_{7/2}}$ levels. 
Laboratory experiments using the EBIT have also been carried out to verify the MIT in Fe~{\sc{x}}. However, large discrepancies exist between the parameters obtained from different approaches. Obviously, more measurements are needed to constrain and improve the current theoretical model and atomic data. 

The suitability of coronal magnetic field measurements using the MIT method has been verified through forward modeling with 3D MHD models of solar and stellar coronae.
It has been shown that accurate measurements of coronal magnetic field using the MIT method require simultaneous observations of the Fe~{\sc{x}} lines in a wide wavelength range of 174--345 {\AA}.
We have also summarized the recent development of the MIT method and its application to magnetic field measurements from Hinode/EIS observations.
The uncertainties and limitations caused by, for example the atomic data, the radiometric calibration, the signal-to-noise ratio, and the limitation of current wavelength coverage from actual solar observations, have also been discussed.
Overall, in order to achieve reliable measurements of the coronal magnetic field using the MIT method, more efforts, particularly more accurate intensity calibration and temperature measurement, need to be made in the future.

\begin{acknowledgements}
This work is supported by NSFC grants 11825301, 12103066, 12073004.
\end{acknowledgements}



\bibliographystyle{raa}
\bibliography{refs}

\begin{thebibliography}{168}
\providecommand\natexlab[1]{#1}
\providecommand\JournalTitle[1]{#1}

\bibitem[{Aggarwal} \& {Keenan}(2005)]{Aggarwal2005}
{Aggarwal}, K.~M., \& {Keenan}, F.~P. 2005, \aap, 439, 1215

\bibitem[{Akhmedov} {et~al.}(1982)]{Akhmedov1982}
{Akhmedov}, S.~B., {Gelfreikh}, G.~B., {Bogod}, V.~M., \& {Korzhavin}, A.~N.
  1982, \solphys, 79, 41

\bibitem[{Akhmedov} {et~al.}(1986)]{Akhmedov1986}
{Akhmedov}, S.~B., {Borovik}, V.~N., {Gelfreikh}, G.~B., {et~al.} 1986, \apj,
  301, 460

\bibitem[{Aly}(1984)]{Aly1984}
{Aly}, J.~J. 1984, \apj, 283, 349

\bibitem[{Andersen} {et~al.}(1993)]{1993PhRvA..47..890A}
{Andersen}, T., {Andersen}, L.~H., {Balling}, P., {et~al.} 1993, \pra, 47, 890

\bibitem[{Anfinogentov} {et~al.}(2019)]{Anfinogentov2019}
{Anfinogentov}, S.~A., {Stupishin}, A.~G., {Mysh'yakov}, I.~I., \& {Fleishman},
  G.~D. 2019, \apjl, 880, L29

\bibitem[{Anzer} \& {Heinzel}(2005)]{Anzer2005}
{Anzer}, U., \& {Heinzel}, P. 2005, \apj, 622, 714

\bibitem[{Argiroffi} {et~al.}(2019)]{2019NatAs...3..742A}
{Argiroffi}, C., {Reale}, F., {Drake}, J.~J., {et~al.} 2019, Nature Astronomy,
  3, 742

\bibitem[{Aschwanden}(2013)]{Aschwanden2013}
{Aschwanden}, M.~J. 2013, \apj, 763, 115

\bibitem[{Aschwanden} {et~al.}(1999)]{Aschwanden1999}
{Aschwanden}, M.~J., {Fletcher}, L., {Schrijver}, C.~J., \& {Alexander}, D.
  1999, \apj, 520, 880

\bibitem[{Badnell}(1997)]{Badnell1997}
{Badnell}, N.~R. 1997, Journal of Physics B Atomic Molecular Physics, 30, 1

\bibitem[{Ballance}(2022)]{Ballance_DARC}
{Ballance}, C.~P. 2022, DARC

\bibitem[{Balling} {et~al.}(1992)]{1992PhRvL..69.1042B}
{Balling}, P., {Andersen}, L.~H., {Andersen}, T., {et~al.} 1992, \prl, 69, 1042

\bibitem[{Banerjee} {et~al.}(2021)]{2021SSRv..217...76B}
{Banerjee}, D., {Krishna Prasad}, S., {Pant}, V., {et~al.} 2021, \ssr, 217, 76

\bibitem[{Bartoe} {et~al.}(1977)]{Bartoe1977}
{Bartoe}, J. D.~F., {Brueckner}, G.~E., {Purcell}, J.~D., \& {Tousey}, R. 1977,
  \ao, 16, 879

\bibitem[{Beiersdorfer} {et~al.}(2004)]{2004RScI...75.3720B}
{Beiersdorfer}, P., {Brown}, G.~V., {Goddard}, R., \& {Wargelin}, B.~J. 2004,
  Review of Scientific Instruments, 75, 3720

\bibitem[{Beiersdorfer} {et~al.}(2016)]{2016ApJ...817...67B}
{Beiersdorfer}, P., {Crespo L{\'o}pez-Urrutia}, J.~R., \& {Tr{\"a}bert}, E.
  2016, \apj, 817, 67

\bibitem[{Beiersdorfer} {et~al.}(2003)]{2003PhRvL..90w5003B}
{Beiersdorfer}, P., {Scofield}, J.~H., \& {Osterheld}, A.~L. 2003, \prl, 90,
  235003

\bibitem[{Berger} {et~al.}(1999)]{Berger1999}
{Berger}, T.~E., {De Pontieu}, B., {Schrijver}, C.~J., \& {Title}, A.~M. 1999,
  \apjl, 519, L97

\bibitem[{Bhatia} \& {Doschek}(1995)]{Bhatia1995}
{Bhatia}, A.~K., \& {Doschek}, G.~A. 1995, Atomic Data and Nuclear Data Tables,
  60, 97

\bibitem[{Brooks} {et~al.}(2021)]{Brooks2021b}
{Brooks}, D.~H., {Warren}, H.~P., \& {Landi}, E. 2021, \apjl, 915, L24

\bibitem[Brooks \& Yardley(2021)]{Brooks2021}
Brooks, D.~H., \& Yardley, S.~L. 2021, Science Advances, 7, eabf0068

\bibitem[{Brosius} {et~al.}(1998)]{Brosius1998}
{Brosius}, J.~W., {Davila}, J.~M., \& {Thomas}, R.~J. 1998, \apjs, 119, 255

\bibitem[{Brown} {et~al.}(2008)]{Brown2008}
{Brown}, C.~M., {Feldman}, U., {Seely}, J.~F., {Korendyke}, C.~M., \& {Hara},
  H. 2008, \apjs, 176, 511

\bibitem[{Chen} {et~al.}(2020)]{Chen2020}
{Chen}, B., {Shen}, C., {Gary}, D.~E., {et~al.} 2020, Nature Astronomy, 4, 1140

\bibitem[{Chen} \& {Peter}(2015)]{2015A&A...581A.137C}
{Chen}, F., \& {Peter}, H. 2015, \aap, 581, A137

\bibitem[{Chen} {et~al.}(2022)]{ChenHechao2022}
{Chen}, H., {Tian}, H., {Li}, H., {et~al.} 2022, \apj, 933, 92

\bibitem[{Chen} {et~al.}(2023)]{Chen2022}
{Chen}, Y., {Bai}, X., {Tian}, H., \& {et al.} 2023, MNRAS

\bibitem[{Chen} {et~al.}(2011)]{Chen2011}
{Chen}, Y., {Feng}, S.~W., {Li}, B., {et~al.} 2011, \apj, 728, 147

\bibitem[{Chen} {et~al.}(2021{\natexlab{a}})]{2021A&A...656L...7C}
{Chen}, Y., {Przybylski}, D., {Peter}, H., {et~al.} 2021{\natexlab{a}}, \aap,
  656, L7

\bibitem[{Chen} {et~al.}(2018)]{2018ApJ...856...21C}
{Chen}, Y., {Tian}, H., {Su}, Y., {et~al.} 2018, \apj, 856, 21

\bibitem[{Chen} {et~al.}(2021{\natexlab{b}})]{Chen2021a}
{Chen}, Y., {Li}, W., {Tian}, H., {et~al.} 2021{\natexlab{b}}, \apj, 920, 116

\bibitem[{Chen} {et~al.}(2021{\natexlab{c}})]{Chen2021b}
{Chen}, Y., {Liu}, X., {Tian}, H., {et~al.} 2021{\natexlab{c}}, ApJL, 918, L13

\bibitem[{Cheng} {et~al.}(2021)]{2021RAA....21..229C}
{Cheng}, G.-C., {Ni}, L., {Chen}, Y.-J., {Ziegler}, U., \& {Lin}, J. 2021,
  Research in Astronomy and Astrophysics, 21, 229

\bibitem[Cheng \& Childs(1985)]{Cheng1985}
Cheng, K.~T., \& Childs, W.~J. 1985, Phys. Rev. A, 31, 2775

\bibitem[{Cheng} {et~al.}(2020)]{2020RAA....20...36C}
{Cheng}, L.-B., {Le}, G.-M., \& {Zhao}, M.-X. 2020, Research in Astronomy and
  Astrophysics, 20, 036

\bibitem[{Cheung} {et~al.}(2019)]{DEM_Cheung}
{Cheung}, M. C.~M., {De Pontieu}, B., {Mart{\'\i}nez-Sykora}, J., {et~al.}
  2019, \apj, 882, 13

\bibitem[{Chifu} {et~al.}(2017)]{Chifu2017}
{Chifu}, I., {Wiegelmann}, T., \& {Inhester}, B. 2017, \apj, 837, 10

\bibitem[{Culhane} {et~al.}(2007)]{Culhane2007}
{Culhane}, J.~L., {Harra}, L.~K., {James}, A.~M., {et~al.} 2007, \solphys, 243,
  19

\bibitem[{De Pontieu} {et~al.}(2009)]{DePontieu2009}
{De Pontieu}, B., {Hansteen}, V.~H., {McIntosh}, S.~W., \& {Patsourakos}, S.
  2009, \apj, 702, 1016

\bibitem[{del Toro Iniesta} \& {Ruiz Cobo}(2016)]{Iniesta2016}
{del Toro Iniesta}, J.~C., \& {Ruiz Cobo}, B. 2016, Living Reviews in Solar
  Physics, 13, 4

\bibitem[{Del Zanna}(2013)]{GDZ}
{Del Zanna}, G. 2013, \aap, 555, A47

\bibitem[{Del Zanna} {et~al.}(2004)]{DelZanna2004}
{Del Zanna}, G., {Berrington}, K.~A., \& {Mason}, H.~E. 2004, \aap, 422, 731

\bibitem[{Del Zanna} {et~al.}(2021)]{chianti10}
{Del Zanna}, G., {Dere}, K.~P., {Young}, P.~R., \& {Landi}, E. 2021, \apj, 909,
  38

\bibitem[{Del Zanna} \& {Mason}(2018)]{DelZanna2018}
{Del Zanna}, G., \& {Mason}, H.~E. 2018, Living Reviews in Solar Physics, 15, 5

\bibitem[{Del Zanna} {et~al.}(2012)]{DelZanna2012}
{Del Zanna}, G., {Storey}, P.~J., {Badnell}, N.~R., \& {Mason}, H.~E. 2012,
  \aap, 541, A90

\bibitem[{Dere} {et~al.}(2019)]{chianti9}
{Dere}, K.~P., {Del Zanna}, G., {Young}, P.~R., {Landi}, E., \& {Sutherland},
  R.~S. 2019, \apjs, 241, 22

\bibitem[{Dere} {et~al.}(1997)]{Dere1997}
{Dere}, K.~P., {Landi}, E., {Mason}, H.~E., {Monsignori Fossi}, B.~C., \&
  {Young}, P.~R. 1997, \aaps, 125, 149

\bibitem[{Donati} {et~al.}(2006)]{2006Sci...311..633D}
{Donati}, J.-F., {Forveille}, T., {Collier Cameron}, A., {et~al.} 2006,
  Science, 311, 633

\bibitem[{Dove} {et~al.}(2011)]{2011ApJ...731L...1D}
{Dove}, J.~B., {Gibson}, S.~E., {Rachmeler}, L.~A., {Tomczyk}, S., \& {Judge},
  P. 2011, \apjl, 731, L1

\bibitem[{Eissner}(1998)]{Eissner1998}
{Eissner}, W. 1998, Computer Physics Communications, 114, 295

\bibitem[{Feldman} {et~al.}(1967)]{1967Phy....33..278F}
{Feldman}, P., {Levitt}, M., {Manson}, S., \& {Novick}, R. 1967, Physica, 33,
  278

\bibitem[{Feng} {et~al.}(2020)]{2020RAA....20..117F}
{Feng}, S., {Deng}, Z., {Yuan}, D., {Xu}, Z., \& {Yang}, X. 2020, Research in
  Astronomy and Astrophysics, 20, 117

\bibitem[{Fleishman} {et~al.}(2020)]{Fleishman2020}
{Fleishman}, G.~D., {Gary}, D.~E., {Chen}, B., {et~al.} 2020, Science, 367, 278

\bibitem[{Froese Fischer} {et~al.}(2019)]{GRASP2018}
{Froese Fischer}, C., {Gaigalas}, G., {J{\"o}nsson}, P., \& {Biero{\'n}}, J.
  2019, Computer Physics Communications, 237, 184

\bibitem[{Froese Fischer} {et~al.}(2016)]{MCDHF}
{Froese Fischer}, C., {Godefroid}, M., {Brage}, T., {J{\"o}nsson}, P., \&
  {Gaigalas}, G. 2016, Journal of Physics B Atomic Molecular Physics, 49,
  182004

\bibitem[{Gan} {et~al.}(2019)]{ASOS}
{Gan}, W.-Q., {Zhu}, C., {Deng}, Y.-Y., {et~al.} 2019, Research in Astronomy
  and Astrophysics, 19, 156

\bibitem[{Gary} \& {Linsky}(1981)]{1981ApJ...250..284G}
{Gary}, D.~E., \& {Linsky}, J.~L. 1981, \apj, 250, 284

\bibitem[{Gary} {et~al.}(2018)]{Gary2018}
{Gary}, D.~E., {Chen}, B., {Dennis}, B.~R., {et~al.} 2018, \apj, 863, 83

\bibitem[{Gibson} {et~al.}(2016)]{2016FrASS...3....8G}
{Gibson}, S., {Kucera}, T., {White}, S., {et~al.} 2016, Frontiers in Astronomy
  and Space Sciences, 3, 8

\bibitem[{Grant} {et~al.}(1980)]{Grant1980}
{Grant}, I.~P., {McKenzie}, B.~J., {Norrington}, P.~H., {Mayers}, D.~F., \&
  {Pyper}, N.~C. 1980, Computer Physics Communications, 21, 207

\bibitem[{Grumer} {et~al.}(2013)]{2013PhRvA..88b2513G}
{Grumer}, J., {Li}, W., {Bernhardt}, D., {et~al.} 2013, \pra, 88, 022513

\bibitem[{G{\"u}del}(2002)]{2002ARA&A..40..217G}
{G{\"u}del}, M. 2002, \araa, 40, 217

\bibitem[{Hale}(1908)]{1908ApJ....28..315H}
{Hale}, G.~E. 1908, \apj, 28, 315

\bibitem[{Hansteen} {et~al.}(2019)]{Hansteen2019}
{Hansteen}, V., {Ortiz}, A., {Archontis}, V., {et~al.} 2019, \aap, 626, A33

\bibitem[{Hou} {et~al.}(2020)]{2020RAA....20...45H}
{Hou}, J.-F., {Xu}, Z., {Yuan}, S., {et~al.} 2020, Research in Astronomy and
  Astrophysics, 20, 045

\bibitem[{Hu} {et~al.}(2022)]{2022SciA....8I9743H}
{Hu}, J., {Airapetian}, V.~S., {Li}, G., {Zank}, G., \& {Jin}, M. 2022, Science
  Advances, 8, eabi9743

\bibitem[{Iwai} \& {Shibasaki}(2013)]{Iwai2013}
{Iwai}, K., \& {Shibasaki}, K. 2013, \pasj, 65, S14

\bibitem[{Jardine} {et~al.}(2002)]{2002MNRAS.333..339J}
{Jardine}, M., {Collier Cameron}, A., \& {Donati}, J.~F. 2002, \mnras, 333, 339

\bibitem[{Ji} {et~al.}(2021)]{2021RAA....21..179J}
{Ji}, H., {Hashim}, P., {Hong}, Z., {et~al.} 2021, Research in Astronomy and
  Astrophysics, 21, 179

\bibitem[{Jiang} {et~al.}(2022)]{Jiang2022}
{Jiang}, C., {Feng}, X., {Guo}, Y., \& {Hu}, Q. 2022, The Innovation, 3, 100236

\bibitem[{Jin} {et~al.}(2020)]{Jin2020}
{Jin}, M., {Cheung}, M.~C.~M., {DeRosa}, M.~L., {et~al.} 2020, in Solar and
  Stellar Magnetic Fields: Origins and Manifestations, ed. A.~{Kosovichev},
  S.~{Strassmeier}, \& M.~{Jardine}, Vol. 354, 426

\bibitem[{Johnstone} {et~al.}(2014)]{2014MNRAS.437.3202J}
{Johnstone}, C.~P., {Jardine}, M., {Gregory}, S.~G., {Donati}, J.~F., \&
  {Hussain}, G. 2014, \mnras, 437, 3202

\bibitem[{Judge} {et~al.}(2016)]{Judge2016}
{Judge}, P.~G., {Hutton}, R., {Li}, W., \& {Brage}, T. 2016, \apj, 833, 185

\bibitem[{Kosugi} {et~al.}(2007)]{Hinode}
{Kosugi}, T., {Matsuzaki}, K., {Sakao}, T., {et~al.} 2007, \solphys, 243, 3

\bibitem[{Kuridze} {et~al.}(2019)]{Kuridze2019}
{Kuridze}, D., {Mathioudakis}, M., {Morgan}, H., {et~al.} 2019, \apj, 874, 126

\bibitem[{Landi} {et~al.}(2006)]{Landi2006}
{Landi}, E., {Del Zanna}, G., {Young}, P.~R., {et~al.} 2006, \apjs, 162, 261

\bibitem[{Landi} {et~al.}(2020{\natexlab{a}})]{Landi2020a}
{Landi}, E., {Hutton}, R., {Brage}, T., \& {Li}, W. 2020{\natexlab{a}}, \apj,
  904, 87

\bibitem[{Landi} {et~al.}(2020{\natexlab{b}})]{Landi2020b}
{Landi}, E., {Hutton}, R., {Brage}, T., \& {Li}, W. 2020{\natexlab{b}}, \apj,
  902, 21

\bibitem[{Landi} {et~al.}(2021)]{Landi2021}
{Landi}, E., {Li}, W., {Brage}, T., \& {Hutton}, R. 2021, \apj, 913, 1

\bibitem[{Lang} \& {Willson}(1986)]{1986ApJ...305..363L}
{Lang}, K.~R., \& {Willson}, R.~F. 1986, \apj, 305, 363

\bibitem[{Li} {et~al.}(2020{\natexlab{a}})]{2020SSRv..216..136L}
{Li}, B., {Antolin}, P., {Guo}, M.~Z., {et~al.} 2020{\natexlab{a}}, \ssr, 216,
  136

\bibitem[{Li} {et~al.}(2018)]{Li2018}
{Li}, D., {Yuan}, D., {Su}, Y.~N., {et~al.} 2018, \aap, 617, A86

\bibitem[{Li} {et~al.}(2017)]{2017ApJ...838...69L}
{Li}, H., {Landi Degl'Innocenti}, E., \& {Qu}, Z. 2017, \apj, 838, 69

\bibitem[{Li} {et~al.}(2013)]{2013PhRvA..88a3416L}
{Li}, J., {Grumer}, J., {Li}, W., {et~al.} 2013, \pra, 88, 013416

\bibitem[{Li}(2022)]{Li2022}
{Li}, W. e.~a. 2022, In preparation

\bibitem[{Li} {et~al.}(2020{\natexlab{b}})]{Li2020}
{Li}, W., {Grumer}, J., {Brage}, T., \& {J{\"o}nsson}, P. 2020{\natexlab{b}},
  Computer Physics Communications, 253, 107211

\bibitem[Li {et~al.}(2021)]{Li_2021}
Li, W., Li, M., Wang, K., {et~al.} 2021, \apj, 913, 135

\bibitem[{Li} {et~al.}(2015)]{Li2015}
{Li}, W., {Grumer}, J., {Yang}, Y., {et~al.} 2015, \apj, 807, 69

\bibitem[{Li} {et~al.}(2016)]{Li2016}
{Li}, W., {Yang}, Y., {Tu}, B., {et~al.} 2016, \apj, 826, 219

\bibitem[{Lin} {et~al.}(2004)]{Lin2004}
{Lin}, H., {Kuhn}, J.~R., \& {Coulter}, R. 2004, \apjl, 613, L177

\bibitem[{Lin} {et~al.}(2000)]{Lin2000}
{Lin}, H., {Penn}, M.~J., \& {Tomczyk}, S. 2000, \apjl, 541, L83

\bibitem[{Liu}(2020)]{2020RAA....20..165L}
{Liu}, R. 2020, Research in Astronomy and Astrophysics, 20, 165

\bibitem[{Liu} {et~al.}(2022)]{Liu2022}
{Liu}, X., {Tian}, H., {Chen}, Y., {et~al.} 2022, \apj, 938, 7

\bibitem[{Liu}(2009)]{2009AnGeo..27.2771L}
{Liu}, Y. 2009, Annales Geophysicae, 27, 2771

\bibitem[{Liu} \& {Lin}(2008)]{2008ApJ...680.1496L}
{Liu}, Y., \& {Lin}, H. 2008, \apj, 680, 1496

\bibitem[{Long} {et~al.}(2017)]{Long2017}
{Long}, D.~M., {Valori}, G., {P{\'e}rez-Su{\'a}rez}, D., {Morton}, R.~J., \&
  {V{\'a}squez}, A.~M. 2017, \aap, 603, A101

\bibitem[{Lu} {et~al.}(2022)]{LuHP2022}
{Lu}, H.-p., {Tian}, H., {Zhang}, L.-y., {et~al.} 2022, \aap, 663, A140

\bibitem[{Madjarska}(2019)]{2019LRSP...16....2M}
{Madjarska}, M.~S. 2019, Living Reviews in Solar Physics, 16, 2

\bibitem[{Magyar} \& {Van Doorsselaere}(2018)]{Magyar2018}
{Magyar}, N., \& {Van Doorsselaere}, T. 2018, \apj, 856, 144

\bibitem[{Malinovsky} {et~al.}(1980)]{Malinovsky1980}
{Malinovsky}, M., {Dubau}, J., \& {Sahal-Brechot}, S. 1980, \apj, 235, 665

\bibitem[{Mannervik} {et~al.}(1997)]{1997HyInt.108..291M}
{Mannervik}, S., {Brostr{\"o}m}, L., {Lidberg}, J., {Norlin}, L.~O., \&
  {Royen}, P. 1997, Hyperfine Interactions, 108, 291

\bibitem[{Mart{\'\i}nez-Sykora} {et~al.}(2022)]{Martinez2022}
{Mart{\'\i}nez-Sykora}, J., {Hansteen}, V.~H., {De Pontieu}, B., \& {Landi}, E.
  2022, \apj, 938, 60

\bibitem[{Mason}(1975)]{Mason1975}
{Mason}, H.~E. 1975, \mnras, 170, 651

\bibitem[{McLaughlin} {et~al.}(2018)]{2018SSRv..214...45M}
{McLaughlin}, J.~A., {Nakariakov}, V.~M., {Dominique}, M., {Jel{\'\i}nek}, P.,
  \& {Takasao}, S. 2018, \ssr, 214, 45

\bibitem[{Mitra-Kraev} {et~al.}(2005)]{2005A&A...436.1041M}
{Mitra-Kraev}, U., {Harra}, L.~K., {Williams}, D.~R., \& {Kraev}, E. 2005,
  \aap, 436, 1041

\bibitem[{Miyawaki} {et~al.}(2016)]{Miyawaki2016}
{Miyawaki}, S., {iwai}, K., {Shibasaki}, K., {Shiota}, D., \& {Nozawa}, S.
  2016, \apj, 818, 8

\bibitem[{Morton} {et~al.}(2015)]{Morton2015}
{Morton}, R.~J., {Tomczyk}, S., \& {Pinto}, R. 2015, Nature Communications, 6,
  7813

\bibitem[{M{\"u}ller} {et~al.}(2020)]{2020A&A...642A...1M}
{M{\"u}ller}, D., {St. Cyr}, O.~C., {Zouganelis}, I., {et~al.} 2020, \aap, 642,
  A1

\bibitem[{Nakajima} {et~al.}(1985)]{NORH1}
{Nakajima}, H., {Sekiguchi}, H., {Sawa}, M., {Kai}, K., \& {Kawashima}, S.
  1985, \pasj, 37, 163

\bibitem[{Nakariakov} \& {Kolotkov}(2020)]{2020ARA&A..58..441N}
{Nakariakov}, V.~M., \& {Kolotkov}, D.~Y. 2020, \araa, 58, 441

\bibitem[{Nakariakov} \& {Ofman}(2001)]{Nakariakov2001}
{Nakariakov}, V.~M., \& {Ofman}, L. 2001, \aap, 372, L53

\bibitem[{Nakariakov} {et~al.}(1999)]{Nakariakov1999}
{Nakariakov}, V.~M., {Ofman}, L., {Deluca}, E.~E., {Roberts}, B., \& {Davila},
  J.~M. 1999, Science, 285, 862

\bibitem[{Ning} {et~al.}(2020)]{2020RAA....20..138N}
{Ning}, Z.-J., {Li}, D., \& {Zhang}, Q.-M. 2020, Research in Astronomy and
  Astrophysics, 20, 138

\bibitem[{Nita} {et~al.}(2016)]{EOVSA}
{Nita}, G.~M., {Hickish}, J., {MacMahon}, D., \& {Gary}, D.~E. 2016, Journal of
  Astronomical Instrumentation, 5, 1641009

\bibitem[{Pandey} \& {Srivastava}(2009)]{2009ApJ...697L.153P}
{Pandey}, J.~C., \& {Srivastava}, A.~K. 2009, \apjl, 697, L153

\bibitem[{Parker}(1983)]{Parker1983}
{Parker}, E.~N. 1983, \apj, 264, 642

\bibitem[{Parker}(1988)]{Parker1988}
{Parker}, E.~N. 1988, \apj, 330, 474

\bibitem[{Pelan} \& {Berrington}(2001)]{Pelan2001}
{Pelan}, J.~C., \& {Berrington}, K.~A. 2001, \aap, 365, 258

\bibitem[{Peter} {et~al.}(2015)]{Peter2015_extraplation}
{Peter}, H., {Warnecke}, J., {Chitta}, L.~P., \& {Cameron}, R.~H. 2015, \aap,
  584, A68

\bibitem[{Peter} {et~al.}(2014)]{2014Sci...346C.315P}
{Peter}, H., {Tian}, H., {Curdt}, W., {et~al.} 2014, Science, 346, 1255726

\bibitem[{Rachmeler} {et~al.}(2013)]{2013SoPh..288..617R}
{Rachmeler}, L.~A., {Gibson}, S.~E., {Dove}, J.~B., {DeVore}, C.~R., \& {Fan},
  Y. 2013, \solphys, 288, 617

\bibitem[{Reiners}(2012)]{2012LRSP....9....1R}
{Reiners}, A. 2012, Living Reviews in Solar Physics, 9, 1

\bibitem[{Roberts} {et~al.}(1984)]{Roberts1984}
{Roberts}, B., {Edwin}, P.~M., \& {Benz}, A.~O. 1984, \apj, 279, 857

\bibitem[{Rodono}(1974)]{1974A&A....32..337R}
{Rodono}, M. 1974, \aap, 32, 337

\bibitem[{Ros{\'e}n} {et~al.}(2015)]{2015ApJ...805..169R}
{Ros{\'e}n}, L., {Kochukhov}, O., \& {Wade}, G.~A. 2015, \apj, 805, 169

\bibitem[{Sandlin} \& {Tousey}(1979)]{Sandlin1979}
{Sandlin}, G.~D., \& {Tousey}, R. 1979, \apjl, 227, L107

\bibitem[{Schatten} {et~al.}(1969)]{Schatten1969}
{Schatten}, K.~H., {Wilcox}, J.~M., \& {Ness}, N.~F. 1969, \solphys, 6, 442

\bibitem[{Schef} {et~al.}(2005)]{2005PhRvA..72b0501S}
{Schef}, P., {Lundin}, P., {Bi{\'e}mont}, E., {et~al.} 2005, \pra, 72, 020501

\bibitem[{Schmieder} {et~al.}(2014)]{Schmieder2014}
{Schmieder}, B., {Tian}, H., {Kucera}, T., {et~al.} 2014, \aap, 569, A85

\bibitem[{Semel}(1989)]{1989A&A...225..456S}
{Semel}, M. 1989, \aap, 225, 456

\bibitem[{Si} {et~al.}(2020)]{Si2020}
{Si}, R., {Brage}, T., {Li}, W., {et~al.} 2020, ApJL, 898, L34

\bibitem[{Soni} {et~al.}(2020)]{2020RAA....20...23S}
{Soni}, S.~L., {Gupta}, R.~S., \& {Verma}, P.~L. 2020, Research in Astronomy
  and Astrophysics, 20, 023

\bibitem[{Sun} {et~al.}(2012)]{Sun2012}
{Sun}, X., {Hoeksema}, J.~T., {Liu}, Y., {et~al.} 2012, \apj, 748, 77

\bibitem[{Tadesse} {et~al.}(2014)]{Tadesse2014}
{Tadesse}, T., {Wiegelmann}, T., {MacNeice}, P.~J., {et~al.} 2014, \solphys,
  289, 831

\bibitem[{Takano} {et~al.}(1997)]{NORH2}
{Takano}, T., {Nakajima}, H., {Enome}, S., {et~al.} 1997, in Coronal Physics
  from Radio and Space Observations, ed. G.~{Trottet}, Vol. 483, 183

\bibitem[{Tan}(2022)]{2022RAA....22g2001T}
{Tan}, B. 2022, Research in Astronomy and Astrophysics, 22, 072001

\bibitem[{Tan} {et~al.}(2016)]{2016RAA....16...82T}
{Tan}, B.-L., {Karlick{\'y}}, M., {M{\'e}sz{\'a}rosov{\'a}}, H., \& {Huang},
  G.-L. 2016, Research in Astronomy and Astrophysics, 16, 82

\bibitem[{Tan} {et~al.}(2020)]{Tan2020}
{Tan}, B.-L., {Yan}, Y., {Li}, T., {Zhang}, Y., \& {Chen}, X.-Y. 2020, Research
  in Astronomy and Astrophysics, 20, 090

\bibitem[{Tayal}(2001)]{Tayal2001}
{Tayal}, S.~S. 2001, \apjs, 132, 117

\bibitem[{Thomas} \& {Neupert}(1994)]{Thomas1994}
{Thomas}, R.~J., \& {Neupert}, W.~M. 1994, \apjs, 91, 461

\bibitem[{Tian} {et~al.}(2012)]{Tian2012}
{Tian}, H., {McIntosh}, S.~W., {Wang}, T., {et~al.} 2012, \apj, 759, 144

\bibitem[{Tian} {et~al.}(2018)]{Tian2018}
{Tian}, H., {Yurchyshyn}, V., {Peter}, H., {et~al.} 2018, \apj, 854, 92

\bibitem[{Tomczyk} {et~al.}(2007)]{Tomczyk2007}
{Tomczyk}, S., {McIntosh}, S.~W., {Keil}, S.~L., {et~al.} 2007, Science, 317,
  1192

\bibitem[{Uchida}(1970)]{Uchida1970}
{Uchida}, Y. 1970, \pasj, 22, 341

\bibitem[{Veronig} {et~al.}(2021)]{2021NatAs...5..697V}
{Veronig}, A.~M., {Odert}, P., {Leitzinger}, M., {et~al.} 2021, Nature
  Astronomy, 5, 697

\bibitem[{Wang} {et~al.}(2015{\natexlab{a}})]{Wang2015nc}
{Wang}, H., {Cao}, W., {Liu}, C., {et~al.} 2015{\natexlab{a}}, Nature
  Communications, 6, 7008

\bibitem[Wang {et~al.}(2020)]{Wang2020}
Wang, K., J\"{o}nsson, P., Zanna, G.~D., {et~al.} 2020, Astrophys. J. Suppl.
  Ser., 246, 1

\bibitem[{Wang} {et~al.}(2012)]{Wang2012}
{Wang}, T., {Ofman}, L., {Davila}, J.~M., \& {Su}, Y. 2012, \apjl, 751, L27

\bibitem[{Wang} {et~al.}(2015{\natexlab{b}})]{Wang2015}
{Wang}, Z., {Gary}, D.~E., {Fleishman}, G.~D., \& {White}, S.~M.
  2015{\natexlab{b}}, \apj, 805, 93

\bibitem[{Wang} {et~al.}(2020)]{2020RAA....20....6W}
{Wang}, Z.-K., {Feng}, S., {Deng}, L.-H., \& {Meng}, Y. 2020, Research in
  Astronomy and Astrophysics, 20, 006

\bibitem[{Warren} {et~al.}(2014)]{HPW}
{Warren}, H.~P., {Ugarte-Urra}, I., \& {Landi}, E. 2014, \apjs, 213, 11

\bibitem[{Welsh} {et~al.}(2006)]{2006A&A...458..921W}
{Welsh}, B.~Y., {Wheatley}, J., {Browne}, S.~E., {et~al.} 2006, \aap, 458, 921

\bibitem[{Wiegelmann} \& {Sakurai}(2021)]{Wiegelmann2021}
{Wiegelmann}, T., \& {Sakurai}, T. 2021, Living Reviews in Solar Physics, 18, 1

\bibitem[{Wiegelmann} {et~al.}(2014)]{Wiegelmann2014}
{Wiegelmann}, T., {Thalmann}, J.~K., \& {Solanki}, S.~K. 2014, \aapr, 22, 78

\bibitem[{Wilhelm} {et~al.}(1995)]{SUMER}
{Wilhelm}, K., {Curdt}, W., {Marsch}, E., {et~al.} 1995, \solphys, 162, 189

\bibitem[{Wilson}(1978)]{Wilson1978}
{Wilson}, O.~C. 1978, \apj, 226, 379

\bibitem[{Wu} {et~al.}(2021)]{2021RAA....21..126W}
{Wu}, W., {Sych}, R., {Chen}, J., \& {Su}, J.-T. 2021, Research in Astronomy
  and Astrophysics, 21, 126

\bibitem[{Xu} {et~al.}(2022)]{Xu2022}
{Xu}, G., {Yan}, C., {Lu}, Q., {et~al.} 2022, \apj, 937, 48

\bibitem[{Yan} {et~al.}(2021)]{MUSER}
{Yan}, Y., {Chen}, Z., {Wang}, W., {et~al.} 2021, Frontiers in Astronomy and
  Space Sciences, 8, 20

\bibitem[{Yang} {et~al.}(2020{\natexlab{a}})]{Yang2020b}
{Yang}, Z., {Tian}, H., {Tomczyk}, S., {et~al.} 2020{\natexlab{a}}, Science
  China Technological Sciences, 63, 2357

\bibitem[{Yang} {et~al.}(2020{\natexlab{b}})]{Yang2020a}
{Yang}, Z., {Bethge}, C., {Tian}, H., {et~al.} 2020{\natexlab{b}}, Science,
  369, 694

\bibitem[{Zhang} {et~al.}(2021)]{2021RAA....21..284Z}
{Zhang}, M.-H., {Zhang}, Y., {Yan}, Y.-H., {et~al.} 2021, Research in Astronomy
  and Astrophysics, 21, 284

\bibitem[{Zhang} {et~al.}(2022)]{2022RAA....22g5007Z}
{Zhang}, X.-F., {Liu}, Y., {Zhao}, M.-Y., {et~al.} 2022, Research in Astronomy
  and Astrophysics, 22, 075007

\bibitem[{Zhao} {et~al.}(2021)]{2021ApJ...912..141Z}
{Zhao}, J., {Gibson}, S.~E., {Fineschi}, S., {et~al.} 2021, \apj, 912, 141

\bibitem[{Zhao} {et~al.}(2019)]{2019ApJ...883...55Z}
{Zhao}, J., {Gibson}, S.~E., {Fineschi}, S., {et~al.} 2019, \apj, 883, 55

\bibitem[{Zhu} {et~al.}(2022)]{Zhu2022}
{Zhu}, X., {Neukirch}, T., \& {Wiegelmann}, T. 2022, Science China
  Technological Sciences, 65, 1710

\bibitem[{Zhu} \& {Wiegelmann}(2018)]{Zhu2018}
{Zhu}, X., \& {Wiegelmann}, T. 2018, \apj, 866, 130

\end{thebibliography}

\end{document}